\documentclass[letterpaper,floatfix,aps,pra,amsmath,amssymb,twocolumn,
showpacs]{revtex4-1}

\usepackage{epsfig}
\usepackage{verbatim}
\usepackage{epsf}
\usepackage{array}
\usepackage{color}
\usepackage[pdfauthor={Giovanni Viola, Gianluigi Catelani},
            pdftitle={Collective modes in the fluxonium qubit}]{hyperref}

\hypersetup{pdfstartview={XYZ null null 1.09}}

\newcommand{\be}{\begin{equation}}
\newcommand{\ee}{\end{equation}}
\newcommand{\bea}{\begin{eqnarray}}
\newcommand{\eea}{\end{eqnarray}}

\newcommand{\eref}[1]{Eq.~(\ref{#1})}
\newcommand{\esref}[1]{Eqs.~(\ref{#1})}
\newcommand{\rref}[1]{(\ref{#1})}

\newcommand{\ocite}[1]{Ref.~\onlinecite{#1}}

\newcommand{\EZO}{\omega_{10}}

\begin{document}

\title{Collective modes in the fluxonium qubit}

\author{G. Viola}

\affiliation{Institute for Quantum Information, RWTH Aachen University, 52056 Aachen, Germany}
\affiliation{Department of Microtechnology and Nanoscience (MC2), Chalmers University of Technology, SE-412 96 Gothenburg, Sweden}

\author{G. Catelani}

\affiliation{Peter Gr\"unberg Institut (PGI-2) and JARA Institute for Quantum Information, Forschungszentrum J\"ulich, 52425 J\"ulich, Germany}

\begin{abstract}
Superconducting qubit designs vary in complexity from single- and few-junction systems, such as the transmon and flux qubits, to the many-junction fluxonium. Here we consider the question of wether the many degrees of freedom in the fluxonium circuit can limit the qubit coherence time. Such a limitation is in principle possible, due to the interactions between the low-energy, highly anharmonic qubit mode and the higher-energy, weakly anharmonic collective modes. We show that so long as the coupling of the collective modes with the external electromagnetic environment is sufficiently weaker than the qubit-environment coupling, the qubit dephasing induced by the collective modes does not significantly contribute to decoherence. Therefore, the increased complexity of the fluxonium qubit does not constitute by itself a major obstacle for its use in quantum computation architectures.
\end{abstract}

\date{\today}

\pacs{74.50.+r, 85.25.Cp}

\maketitle

\section{Introduction}
\label{sec:intro}

Starting from the pioneering experiments with Cooper pair boxes \cite{cpb}, superconducting qubits have been vastly improved \cite{science}, thanks in part to better control of the qubit environment. For example, 3D transmons \cite{3dtr} benefit from being placed in a cavity that suppress radiative losses as well as from their relatively large size that decreases the role of surface dielectric losses. Planar transmon variants such as the Xmon \cite{xmon} have shorter coherence times, but have made possible demonstrations of the building blocks of quantum error correction codes \cite{ecc1,ecc2,ecc3,ecc4}. In contrast to the transmon architecture, in which a Josephson junction is shunted capacitively, in a fluxonium qubit the shunt is via an inductance \cite{flux1,flux2}. To realize in practice a sufficiently large inductance, arrays with many Josephson junctions are used. Even in this more complicated circuit, long relaxation times have been achieved -- so long, in fact, to make possible the accurate measurement of the phase dependence of the quasiparticle dissipation through a Josephson junction \cite{nature}. Experimentally, the fluxonium coherence time is much shorter than the limit imposed by relaxation; in this paper we investigate whether the complexity of the circuit, arising from the use of junction arrays, contributes to this shortness.

The physics of arrays made of identical Josephson junctions has long attracted the interest of theorists and experimentalists alike, especially in the context of the superconductor-insulator transition controlled by the ratio of charging and Josephson energies \cite{sit1,sit2}. More recently, renewed attention has been given to the physics of collective modes, whose frequency can be significantly lowered below the constituent Josephson junctions plasma frequency due to ground capacitances; in particular, the interplay between collective modes and quantum phase slips in rings has been studied \cite{qps1,qps2,qps3}, as well as the localizing effect of disorder in the junction parameters \cite{disorder}. Long junction arrays can have a large inductance, proportional to the number of elements, and for quantum computation applications such superinductors \cite{supind} have been proposed as elements of topologically protected qubits \cite{toprev}, for example the $0$-$\pi$ qubit \cite{zp1,zp2}.
In the fluxonium, even in the ideal case of no capacitances to ground, the full symmetry of a ring is broken by the presence of a smaller junction. Nonetheless, in the ideal case a permutation symmetry is still present; the effect of this symmetry and its breaking on the fluxonium spectrum has been studied in \ocite{prx}. Aspects of the quasiparticle-induced decoherence in the fluxonium have been studied theoretically in \cite{prb1,prb2,fj}. Here we focus on possible limitation of coherence of the qubit mode due to its interaction with the collective modes of the array. We will show that both ground capacitances and array-junction non-linearities introduce potential decoherence channels, but that they are not limiting the fluxonium coherence. We nonetheless point out that isolation of the collective modes from the electromagnetic environment is necessary for long coherence times.

The paper is organized as follows: in Sec.~\ref{sec:model} we introduce our model for the fluxonium circuit; it includes ground and coupling capacitors in addition to the charging and Josephson energies of each junction. Section~\ref{sec:odd} discusses the properties of the odd collective modes, which are decoupled from the qubit in the parity and time reversal ($PT$-)symmetric case. The interactions between even and qubit modes due to ground capacitances is the subject of Sec.~\ref{sec:even}. In Sec.~\ref{sec:anhar} we consider the effects of the array-junction non-linearity.
In Sec.~\ref{sec:coher}, using the results of the previous two sections we estimate the fluxonium dephasing rate due to the interactions with the collective modes.
The consequences of breaking $PT$ symmetry and of placing the coupling capacitors in the array are briefly discussed in Sec.~\ref{sec:sb} and \ref{sec:sic}, respectively.
We summarize our main results in Sec.~\ref{sec:summ}. Numerous technical details are presented in Appendices~\ref{app:lder} through \ref{app:hcs}. Throughout the paper we set
$\hbar = 1$.

\section{Fluxonium model}
\label{sec:model}

\begin{figure}[b]
 \includegraphics[width=0.48\textwidth]{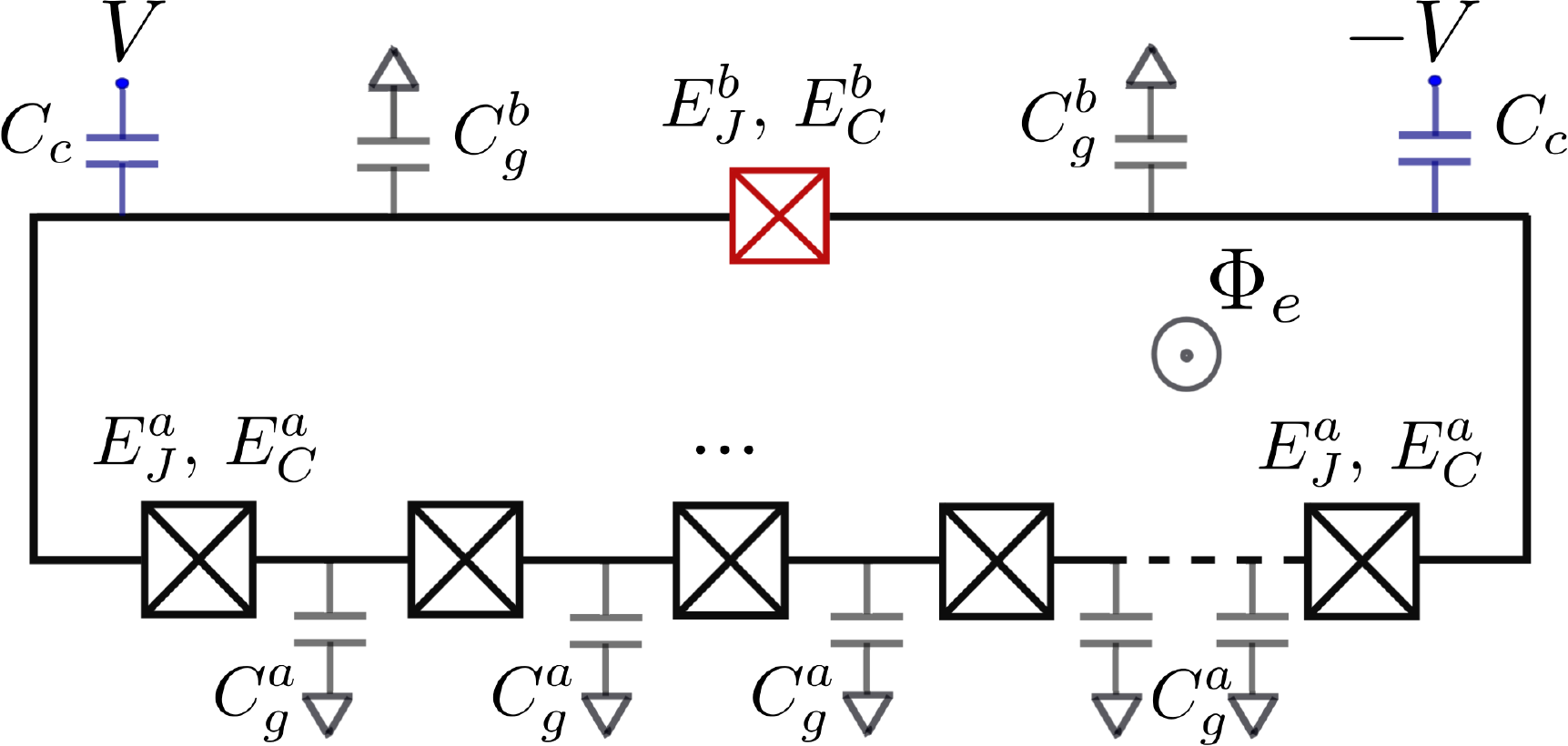}
 \caption{Circuit model for the fluxonium: a small junction (top, Josephson energy $E_J^b$, charging energy $E_C^b$) shunts an array (bottom) of many identical junctions. A capacitance ($C_g^a$ or $C_g^b$) is present between each superconducting island and ground. Additional capacitors $C_c$, biased at voltages $\pm V$, are used to control and read out the qubit, whose spectrum depends on the applied flux $\Phi_e$.}
 \label{flu1_b}
\end{figure}

The circuit model for the fluxonium qubit is shown in Fig.\ref{flu1_b}: the two ends of an array of $N\gg1$ identical Josephson junctions (Josephson energy $E_J^a$ and charging energy $E_C^a=e^2/2C_J^a$) are connected by a so-called phase slip junction (Josephson energy $E_J^b$ and charging energy $E_C^b$). The loop thus formed is pierced by a magnetic flux $\Phi_e$. To suppress phase fluctuations in the array, we require $E_J^a/E_C^a \gg 1$; phase slips preferentially take place at the phase-slip junction so long as $E_J^b/E_C^b < E_J^a/E_C^a$. The superconducting islands inside the array have ground capacitances $C_g^a$ and those at the ends $C_g^b$. Identical coupling capacitors of capacitance $C_c$, used to control the system by applying ac voltages $+V$ and $-V$ to them, are connected to the islands at the end of the array. The Lagrangian description of this circuit without the coupling capacitors is discussed in detail in \ocite{prx}. In Appendix~\ref{app:lder} we briefly summarize (for the paper to be self-contained) and generalize (to include the coupling capacitors) the relevant parts of that work.

While there are $N+1$ junctions in the circuit, due to charge
conservation and flux quantization there are only $N$ independent degrees of freedom \cite{prx}: the qubit mode $\phi$ and the collective modes $\xi_\mu$, $\mu = 1, \ldots, N-1$. In terms of these modes, the Lagrangian ${\cal L}_S$ in the absence of ground and coupling capacitors takes the form
\bea
{\cal L}_S &=& T_S - U_S \, , \\
T_S & = & \frac{1}{16 E_C^\phi} \dot\phi^2 + \frac{1}{16 E_C^a} \sum_\mu \dot\xi_\mu^2 \, , \label{TS} \\
U_S & = &  - E_J^b \cos \left(\phi + \varphi_e \right) \label{US}  \\ && - E_J^a \sum_m \cos \left[ \frac{\phi}{N} + \sum_\mu W_{\mu m} \xi_\mu \right] \, , \nonumber
\eea
where the phase bias $\varphi_e$ is due to the externally applied magnetic flux, $\varphi_e = 2\pi \Phi_e/\Phi_0$ with $\Phi_0$ the flux quantum, and the qubit mode charging energy $E_C^\phi$ is
\be
\frac{1}{E_C^\phi} = \frac{1}{E_C^b} + \frac{1}{N E_C^a} \, ;
\ee
note that due to the last term in the above equation, $E_C^\phi$ is smaller than the phase slip junction charging energy $E_C^b$.
Hereinafter, sums over index $m$ run from $1$ to $N$ and those over Greek indices such as $\mu$ from $1$ to $N-1$. To write the Lagrangian in the given form, the matrix $W_{\mu m}$ must satisfy $\sum_m W_{\mu m}W_{\nu m} = \delta_{\mu\nu}$ and $\sum_m W_{\mu m} = 0$; for concrete calculations, we will use for $W_{\mu m}$ the form suggested in \ocite{prx}:
\be\label{wmn}
W_{\mu m} = \sqrt{\frac{2}{N}} \, \cos \frac{\pi \mu (m-1/2)}{N} \, .
\ee
It can be shown \cite{prx} that ${\cal L}_S$ is symmetric under the action of the symmetric group $\mathrm{S}_N$.

Due to the large ratio $E_J^a/E_C^a$, fluctuations of $\xi_\mu$ are small; for fluctuations in $\phi$ small compared to $\pi N$, we can then expand the last term in $U_S$ to quadratic order to find (up to a constant term)
\be\label{U}
U_S \simeq U = -E_J^b \cos \left(\phi+ \varphi_e \right) + \frac12 E_L \phi^2 + \frac12 E_J^a \sum_\mu \xi_\mu^2
\ee
with $E_L = E_J^a/N$. At this lowest order, the qubit mode $\phi$ and collective modes $\xi_\mu$ do not interact (so long as we neglect ground and coupling capacitors), and
the Lagrangian
\be\label{LU}
{\cal L}_U = T_S - U
\ee
is symmetric under the unitary group $\mathrm{U}(N-1)$. We note that the symmetry under permutations ensures \cite{prx} that the anharmonic terms that we neglect cannot couple any state given by the direct product by a qubit eigenstate and a collective modes singly excited state to a state which is the direct product between any qubit state and the collective modes ground state. One of the main objectives of this work is to understand the interactions induced by the presence of ground and coupling capacitors; we will show, for example, that the coupling between the states just discussed is in general present, albeit weak.

Ground and coupling capacitors modify the kinetic energy part of the Lagrangian by the addition of the term $T_G$ given by
\be\label{TG}
T_G = \frac{1}{16} \left[ G_{00} \dot\phi^2 + 2 \sum_\mu G_{0\mu} \dot\phi \dot\xi_\mu + \sum_{\mu\nu} G_{\mu\nu} \dot\xi_\mu \dot\xi_\nu \right],
\ee
where the symmetric matrix $G$ has entries
\bea
G_{00} &= &\frac{1}{4 E_t}\left[1 - \frac23 \frac{N+1}{N} \lambda \right]\, ,  \label{G00} \\
G_{0\mu} & = & - \frac{1}{2E_g^a}\frac{c_\mu o_{\mu+1}}{\sqrt{2N} s_\mu^2} \, ,\label{G0m} \\
G_{\mu\nu} & = & \frac{1}{4E_g^a} \frac{1}{s_\mu^2} \left[\delta_{\mu\nu} - \lambda \frac{2}{N(N-1)} \frac{o_\mu o_\nu c_\mu c_\nu}{s_\nu^2} \right].
\eea
Here the energy scales are $E_g^a = e^2/2C_g^a$ and $E_t = e^2/2C_t$, where the total capacitance $C_t$ is
\be\label{Ctdef}
C_t = 2 \left(C_g^b + C_c\right) + (N-1) C_g^a \, .
\ee
The dimensionless parameter $0\le \lambda \le 1$ is defined as
\be\label{lam_def}
\lambda = \frac{(N-1)C_g^a}{C_t}
\ee
and we introduced the short-hand notations
\bea
s_\mu &=& \sin \frac{\pi \mu}{2N} \, ,\label{sm} \\
c_\mu &=& \cos \frac{\pi \mu}{2N} \, , \\
o_\mu &=& \frac{1-(-1)^\mu}{2} \, .\label{om}
\eea

The coupling capacitors enable control of the circuit via external ac voltages; the coupling Lagrangian ${\cal L}_V$ takes the form
\be\label{LV1}
{\cal L}_V = - \frac{1}{4 E_C^c} \dot\phi \, e V \, ,
\ee
where we assumed that the voltages applied to the capacitors are equal in magnitude $V$ and opposite in sign and $E_C^c = e^2/2C_c$.

From the above definitions, it is evident that the collective modes with even index $\mu$ interact only with the qubit mode, whereas the odd modes interact only among themselves. Therefore the total approximate \cite{lappr} Lagrangian ${\cal L}$ separates into a sum of even and odd sectors:
\be\label{Ltot}
{\cal L} = {\cal L}_U + T_G + {\cal L}_V = {\cal L}_e + {\cal L}_o \, .
\ee
This separation is valid as long as parity $P$ and time reversal $T$ symmetries are preserved \cite{prx}: indeed, the qubit mode and the collective modes with even index $\mu$ are even under $PT$ symmetry, while modes with odd index are odd. We restrict our attention to the $PT$-symmetric case for most of the paper, but we discuss the consequences of breaking this symmetry in Sec.~\ref{sec:sb}. In the next section we focus on the odd sector. Throughout the paper we will present examples calculated with the two parameter set given in Table~\ref{tabpar}; as explained in Appendix~\ref{app:par}, the parameters are chosen as to reflect realistic experimental values \cite{flux1,nature}.

\begin{table}[b]
\begin{tabular}{|c||c|c|c|c|c|c|c|c||c|c|}
\hline
& $N$ & $E_J^a$ & $E_C^a$ & $E_J^b$ & $E_C^b$ & $E_g^a$ & $E_g^b$ & $E_C^c$ & $\lambda$ & $N_s$ \\ \hline
set 1 & 43 & 26.0 & 1.24 & 8.93 & 3.60 & 194 & 6 & 24.2 & 0.34 & 39\\
set 2 & 95 & 48.3 & 1.01 & 10.2 & 4.78 & 484 & 807 & 12.1 & 0.54 & 69 \\ \hline
\end{tabular}
\caption{Fluxonium parameters used for numerical calculations throughout the paper (see also Appendix~\ref{app:par}). Energies are given in GHz, while the number of junctions $N$, $\lambda$ [\eref{lam_def}], and the screening length $N_s$ [\eref{smallN}] are dimensionless.}
\label{tabpar}
\end{table}

\section{Collective modes: odd sector}
\label{sec:odd}

The collective modes odd under $PT$-symmetry are governed by the Lagrangian
\be\label{Lo}
{\cal L}_o = T_o - \frac{1}{2} E_J^a \sum_{\rho=1}^{N_o} \zeta^2_{\rho} \, ,
\ee
where to simplify the notation we introduce $\zeta_\rho = \xi_{2\rho-1}$ and the number of odd modes $N_o = \lfloor N/2 \rfloor$ equal to the integer part of $N/2$. In the kinetic energy term, we separate a purely diagonal term $T_d$ independent of $\lambda$ , and a term $T_\lambda$ proportional to $\lambda$:
\bea
T_o &=& T_d + \lambda T_\lambda \, , \label{To0} \\
T_d &=& \frac{1}{16} \sum_{\rho=1}^{N_o} \left[\frac{1}{E_C^a} + \frac{1}{4E_g^a}\frac{1}{s^2_{2\rho-1}}\right]\dot\zeta_\rho^2 \, ,\\
T_\lambda &=& -\frac{1}{32E_g^a}\sum_{\sigma,\rho=1}^{N_o} \frac{1}{N(N-1)}\frac{c_{2\sigma-1}c_{2\rho-1}}{s^2_{2\sigma-1}s^2_{2\rho-1}} \dot\zeta_\sigma \dot\zeta_\rho \, .
\eea

If the number of junctions is not too large,
\be\label{smallN}
N \ll N_s \equiv \pi \left(\frac{E_g^a}{E_C^a}\right)^{1/2} ,
\ee
the effect of the ground capacitances can be treated perturbatively for all modes and all values of $\lambda$;
if the array is short compared to the ``screening length'' $N_s$, the ground capacitances hardly affect the energy spectrum of the modes \cite{qps1}.
The condition \rref{smallN} is not satisfied in current experiments, see Table~\ref{tabpar}. However, as the order number $\sigma$ of the modes increases, the effect rapidly diminishes \cite{fpert}; moreover, the off-diagonal part is proportional to $\lambda$, which is typically somewhat smaller than unity. This points to the more general viability of a perturbative approach  than suggested by the condition \rref{smallN} above.
Formally, diagonalization of the kinetic energy matrix can be obtained by a rotation that eliminates the off-diagonal terms [see Appendix~\ref{app:rot}; note that since the potential energy term in \eref{Lo} is quadratic and proportional to the identity matrix, any rotation leaves it unchanged]; up to second order in $\lambda$  we find:
\be
T_o \simeq \frac{1}{16}\sum_{\rho=1}^{N_o} \frac{1}{E_{C,\rho}^o} \dot{\tilde\zeta}_\rho^2 \, ,
\ee
where the effective charging energy for the new, rotated modes $\tilde\zeta_\rho$ is
\be\label{ecodd}\begin{split}
& \frac{1}{E_{C,\rho}^o} \simeq \frac{1}{E_C^a} + \frac{1}{4E_g^a} \bigg[\frac{1}{s_{2\rho-1}^2}-\lambda \frac{2}{N(N-1)} \frac{c_{2\rho-1}^2}{s_{2\rho-1}^4} \\
& \quad + \left(\lambda \frac{2}{N(N-1)}\right)^2\sum_{\sigma\neq\rho}\frac{c^2_{2\rho-1} c^2_{2\sigma-1}}
{s^2_{2\rho-1} s^4_{2\sigma-1} - s^4_{2\rho-1} s^2_{2\sigma-1}} \bigg].
\end{split}\ee
Comparison with numerical diagonalization for a range of experimentally relevant parameters shows that this formula is accurate to a few percent even for the lower energy modes (and much better accuracy for the high-energy modes), and that the term proportional to $\lambda^2$ does not contribute significantly to the effective charging energy, thus validating our perturbative approach. With the Lagrangian in diagonal form, it is straightforward to obtain the energy spectrum of the modes:
\be
\omega_\rho^o = \sqrt{8 E_{C,\rho}^o E_J^a} \ .
\ee
In Sec.~\ref{sec:expcomp} we will briefly compare the calculated spectrum with recent experiments \cite{supind,wg}. In the next section we turn our attention to the even sector.

\section{Even sector and qubit mode}
\label{sec:even}

The qubit mode $\phi$ belongs to the even sector, and in the presence of capacitance to ground in the array it interacts with all the even modes:
\be\label{Leven}\begin{split}
{\cal L}_e = & \, {\cal L}_\phi + T_e - \frac{1}{2} E_J^a \sum_{\rho=1}^{N_e} \eta^2_{\rho} +{\cal L}_V \\
& \, - \frac{1}{16\sqrt{2N}}\frac{1}{E_g^a} \, \dot\phi \sum_{\rho=1}^{N_e} \frac{c_{2\rho}}{s^2_{2\rho}} \dot\eta_\rho \, .
\end{split}\ee
Here we define $\eta_\rho = \xi_{2\rho}$ and the number of even modes is $N_e= \lfloor (N-1)/2 \rfloor$. The kinetic energy of the even modes has a simple diagonal form
\be\label{eceven}
T_e = \frac{1}{16} \sum_{\rho=1}^{N_e} \left[\frac{1}{E_C^a} + \frac{1}{4E_g^a} \frac{1}{s^2_{2\rho}} \right] \dot\eta^2_{\rho} \equiv \frac{1}{16} \sum_{\rho=1}^{N_e} \frac{1}{E_{C,\rho}^e}  \dot\eta^2_{\rho}
\ee
and the qubit Lagrangian is
\be\label{Lphi}
{\cal L}_\phi =  \frac{1}{16\tilde{E}_C^\phi} \dot\phi^2
 + E_J^b \cos\left(\phi + \varphi_e \right) -\frac12 E_L \phi^2 \, ,
\ee
where the effective qubit charging energy is given by
\be\label{tEcp}
\frac{1}{\tilde{E}_C^\phi} = \frac{1}{E_C^\phi} + \frac{1}{4E_t} \left(1-\frac23 \frac{N+1}{N} \lambda \right) .
\ee

\begin{figure}[tb]
 \includegraphics[width=0.48\textwidth]{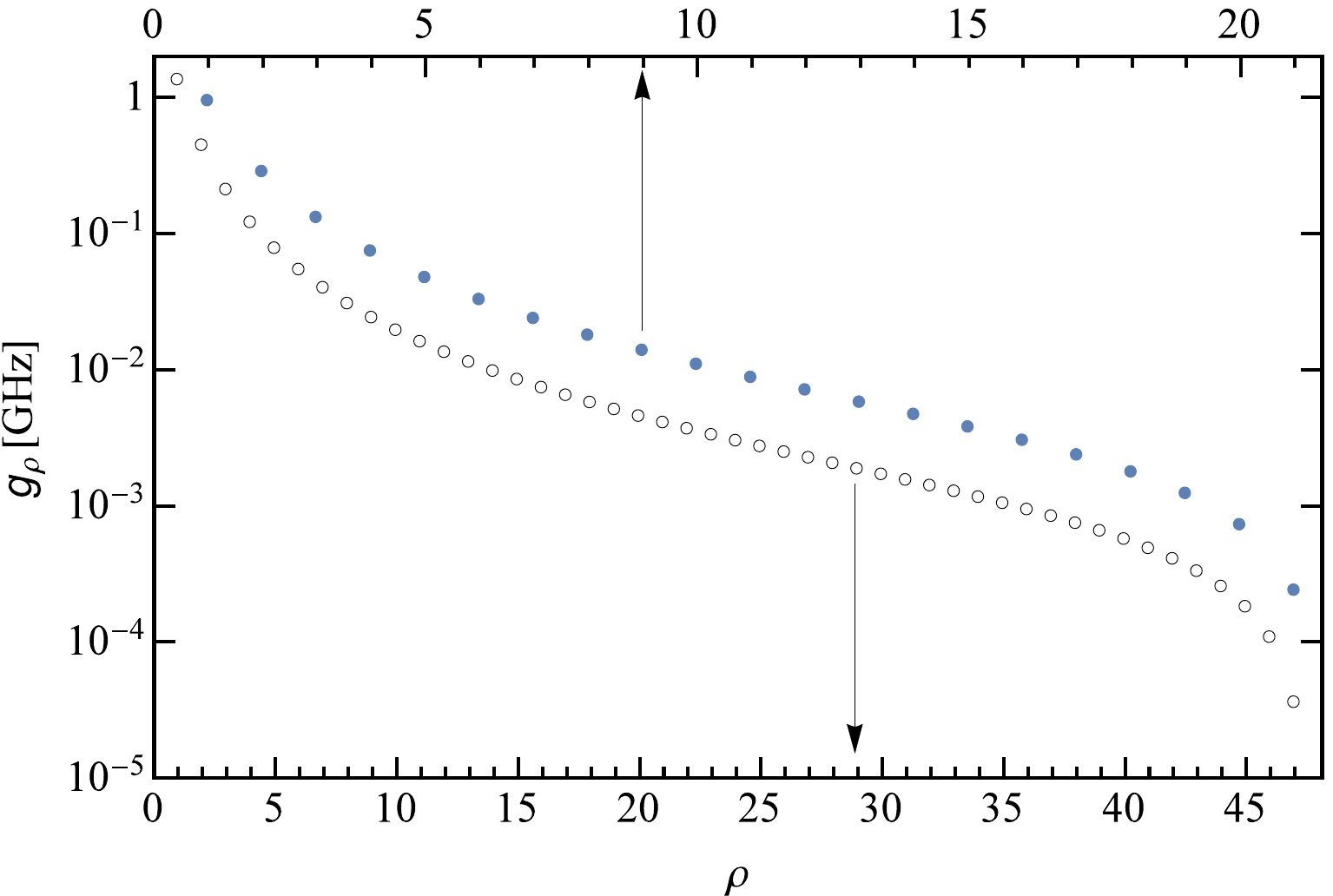}
 \caption{(Color online) Qubit-collective modes coupling constants $g_\rho$, \eref{Hint}, calculated using parameter set 1 in Table~\ref{tabpar} (filled circles) and 2 (empty cricles). Different horizontal scales are used for the two set, as indicated by the arrows, and hence $g_\rho$ for set 2 is larger than the corresponding coupling constant for set 1. Note that in absolute value the coupling strength between qubit and the few lowest even modes is as strong as or even stronger than the typical qubit-cavity coupling \cite{flux1,flux2}.}
 \label{grfig}
\end{figure}

The Lagrangian ${\cal L}_e$ has a simple structure, describing a set of independent harmonic oscillators interacting with the qubit mode. To this Lagrangian, however,
corresponds an Hamiltonian $H_e$ in which all degrees of freedom interact among themselves [see Appendix~\ref{app:He}] due the non diagonal form of the kinetic energy -- cf. the last term in \eref{Leven}. The condition in \eref{smallN} is again sufficient to enable perturbative treatment of these interactions but, as mentioned above, it is not experimentally satisfied. In contrast, the generally weaker condition
\be\label{Hcon}
N \ll 8\pi^2 \frac{E_g^a}{\tilde{E}_C^\phi}
\ee
is satisfied in current experiments, with the right hand side being about 6400 (10600) for parameter set 1 (2). This conditions enables us to
make substantial simplifications, with the approximate Hamiltonian taking the form
\bea
H_e & = & H_\phi + \sum_{\rho=1}^{N_e} H_\rho + H_\mathrm{int} + H_V \, ,\label{He} \\
H_\phi & = & 4\tilde{E}_C^\phi p_\phi^2 - E_J^b \cos \left(\phi + \varphi_e\right) + \frac12 E_L \phi^2 \, , \label{Hphi} \\
H_\rho & = & 4E_{C,\rho}^e p_\rho^2 +\frac12 E_J^a \eta_\rho^2 \, ,\\
H_\mathrm{int} & = & \sum_{\rho=1}^{N_e} g_\rho p_\rho p_\phi \, , \quad g_\rho = \frac{4}{\sqrt{2N}} \frac{\tilde{E}_C^\phi E_{C,\rho}^e}{E_g^a} \frac{c_{2\rho}}{s_{2\rho}^2} \, , \label{Hint} \\
H_V & = & -2 \frac{\tilde{E}_C^\phi}{E_C^c} p_\phi \, eV -  \sum_{\rho=1}^{N_e} \frac{1}{4E_C^c} g_\rho p_\rho \,eV \, .\label{HV}
\eea
Here $p_\rho$ is the momentum conjugate to $\eta_\rho$.
The above expression for the Hamiltonian is one of the main results of this paper: it contains the leading interaction terms between the qubit mode and the collective modes of the junctions forming the superinductance due to capacitance to ground in the array. The coupling constant $g_\rho$ is proportional to the array capacitance to ground and monotonically decreases with $\rho$, see Fig.~\ref{grfig}; therefore the higher collective modes couple more weakly to the qubit. Similarly, since $g_\rho$ determines also the coupling of the collective modes to the ac voltage $V$, the higher modes are more weakly coupled to $V$ than the lower ones; moreover, the low-energy ones are more weakly coupled to $V$ than the qubit mode. As we will see below, this implies a lower decay rate for the collective modes compared to the qubit.

\subsection{Dispersive shifts}

To further study the effect on the qubit of the interaction with the collective modes, we
perform a Schrieffer-Wolff transformation and project the Hamiltonian $H_e$, \eref{He}, onto the qubit subspace to find the effective Hamiltonian \cite{fluxth}
\be\label{Heff}
H_\mathrm{eff} = \frac{\omega_{10}(f)}{2} \sigma^z + \sum_{\rho=1}^{N_e}  \left[\omega_\rho +\chi_\rho(f) \sigma^z \right] a^\dagger_\rho a_\rho \, ,
\ee
where $\sigma^z$ is a Pauli matrix in the qubit subspace,
\be
\omega_\rho^e = \sqrt{8 E_{C,\rho}^e E_J^a}
\ee
is the harmonic oscillator frequency of the even collective mode $\rho$, and $a^\dagger_\rho$, $a_\rho$ the creation and annihilation operators. As indicated by the presence of the parameter $f = \varphi_e/2\pi$, the qubit frequency $\omega_{10}$ and the ac Stark shifts $\chi_\rho$ depend on flux through the loop, and we neglect for the moment the coupling to external bias given by $H_V$. Here with the frequency $\omega_{10}$ we indicate the energy difference between the two lowest eigenstates of the Hamiltonian $H_\phi$, \eref{Hphi}; below we will discuss a small renormalization $\delta\omega_{10}$ of the qubit frequency due to the interaction with the collective modes.

\begin{figure}[bt]
 \includegraphics[width=0.48\textwidth]{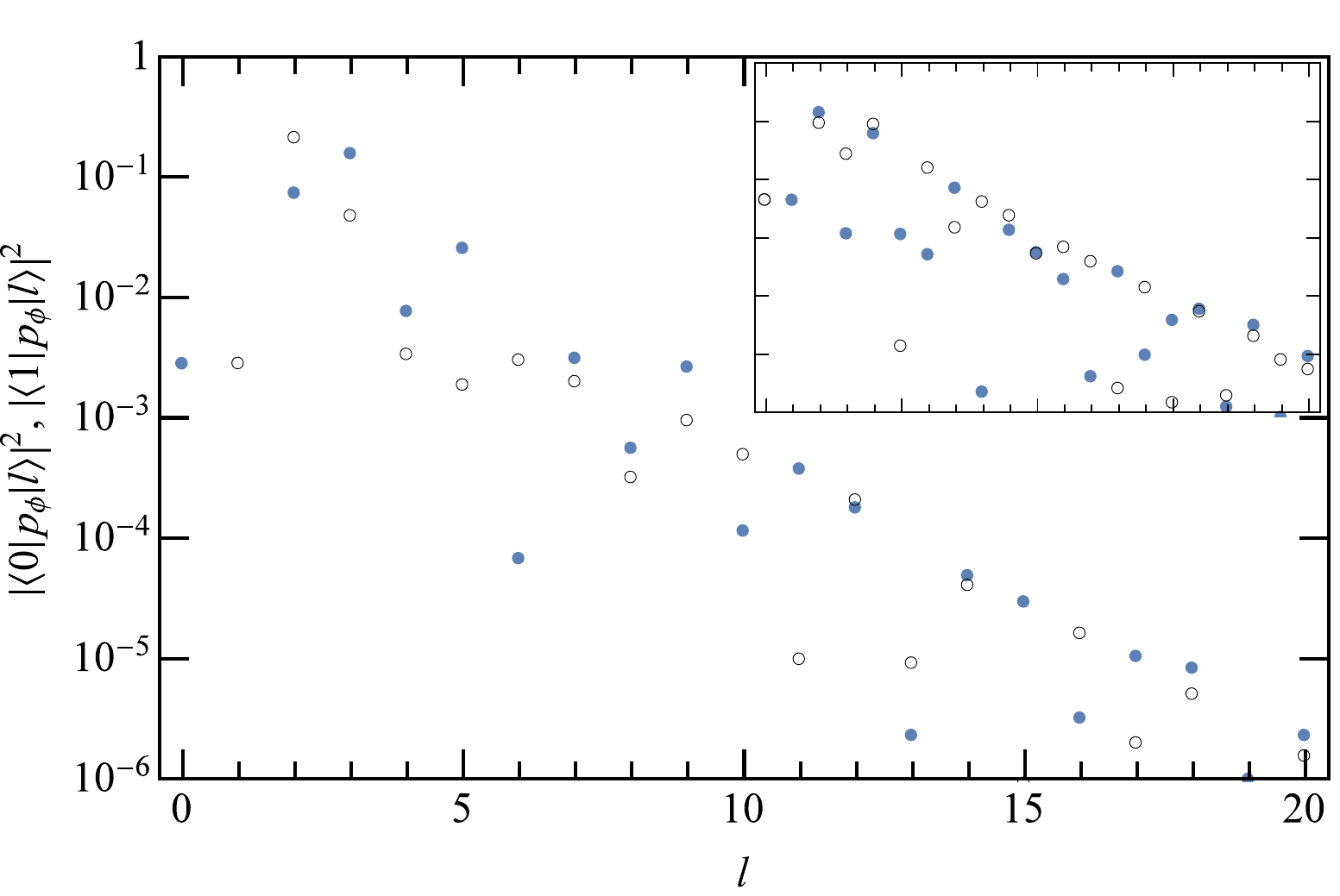}
 \caption{(Color online) Main panel: matrix elements squared $|\langle 0 | p_\phi | l \rangle|^2$ (filled circles) and  $|\langle 1 | p_\phi | l \rangle|^2$ (empty circles) at $f=0.35$ for $l\le 20$ calculated using parameter set 1 [see Table~\ref{tabpar}]. Inset: same as main panel but for parameter set 2. We stress that the overall decrease of the matrix elements with $l$ is valid for any flux $f$, not just for the particular value used here as an example.}
 \label{pme}
\end{figure}

The dispersive shifts $\chi_\rho$ depend on matrix elements of charge operator $p_\phi$ that involve all the eigenstates $|l\rangle$ with energy $\epsilon_l$ ($l=0,1,2,\ldots$) of Hamiltonian  $H_\phi$ \cite{fluxth}:
\be\label{chir}\begin{split}
\chi_\rho = \frac12 \sqrt{\frac{E_J^a}{8E^e_{C,\rho}}} \, g_\rho^2 \bigg[ & \left|\langle 0 | p_\phi | 1 \rangle\right|^2 \frac{2\omega_{10}}{\omega_{10}^2 - \omega_\rho^2} \\
+ \sum_{l\ge 2} & \left|\langle 0 | p_\phi | l \rangle\right|^2 \frac{\omega_{l0}}{\omega_{l0}^2 - \omega_\rho^2} \\
- \sum_{l\ge 2} & \left|\langle 1 | p_\phi | l \rangle\right|^2 \frac{\omega_{l1}}{\omega_{l1}^2 - \omega_\rho^2} \bigg]
\end{split}\ee
with $\omega_{lj} = \epsilon_l - \epsilon_j$.
While in the case of the transmon \cite{pratr} selection rules and low anharmonicity enable the analytical calculation of the dispersive shift, for the fluxonium this is not possible, so we resort to numerical estimates. Fast convergence of the sums in the above equation is ensured by the rapid decrease of the matrix elements $\langle 0 | p_\phi | l \rangle$ and $\langle 1 | p_\phi | l \rangle$ as $l$ increases, see Fig.~\ref{pme}; further aiding the convergence is the approximately linear increase of the energy of the states with slope $\sim(8 \tilde{E}_C^\phi E_L)^{1/2}$. The reason for the decay of the matrix elements is the following: the low-lying states are localized near $\phi_e$, while the high-energy states are to a good approximation the eigenstates of the harmonic oscillator obtained by neglecting the Josephson term in $H_\phi$ (this also explain the linear increase in their energy). Since these high-energy states display oscillations of small magnitude at the center of the potential, the overlap of their derivative with the low-lying states is small, and the increase of the number of oscillations with $l$ causes the decrease of the matrix elements.

\begin{figure}[bt]
 \includegraphics[width=0.48\textwidth]{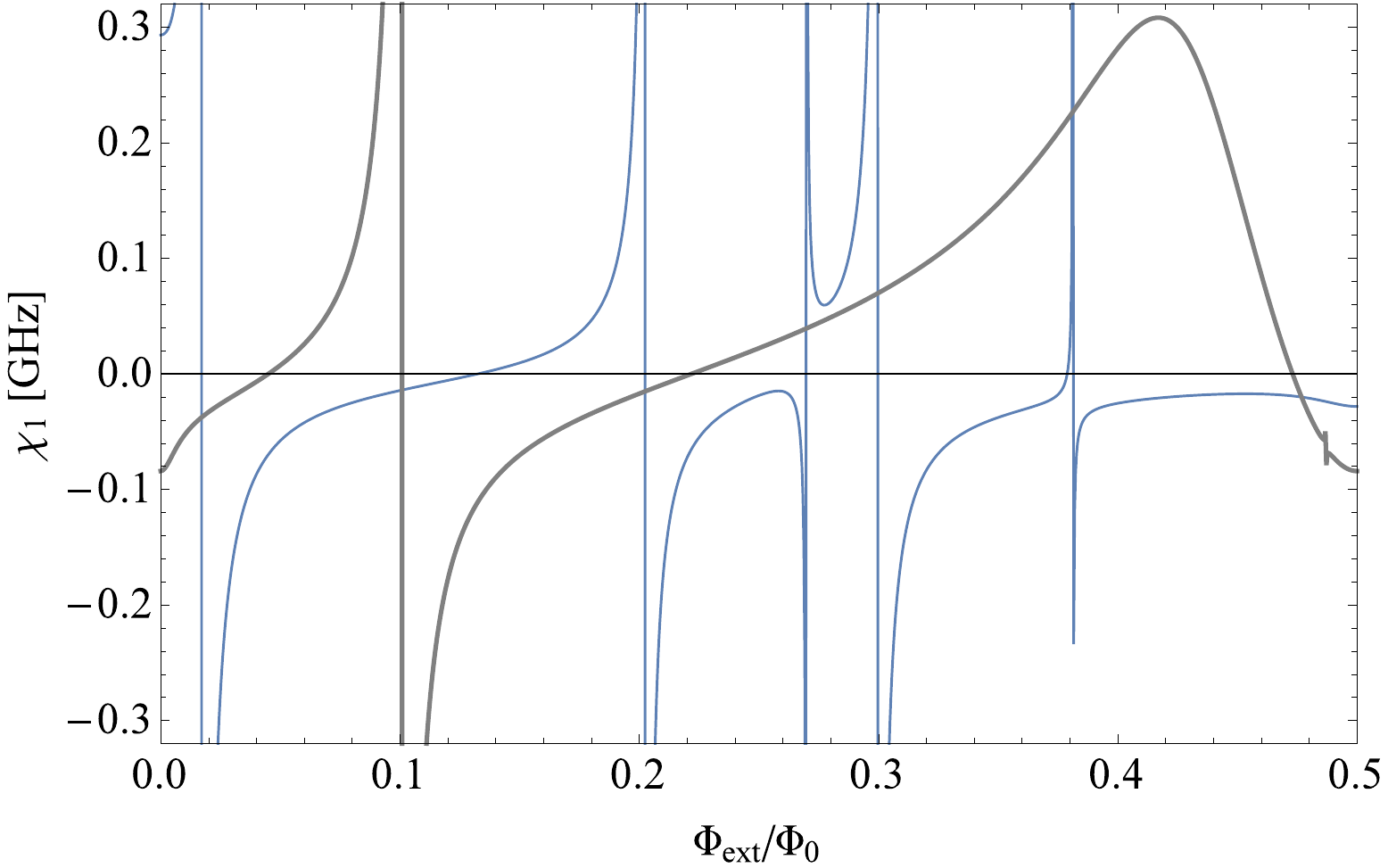}
 \caption{(Color online) Dispersive shift $\chi_1$ [\eref{chir}] as function of external flux for parameter set 1 (thin line) and 2 (thick line).}
 \label{chi1fig}
\end{figure}

\begin{figure}[bt]
 \includegraphics[width=0.48\textwidth]{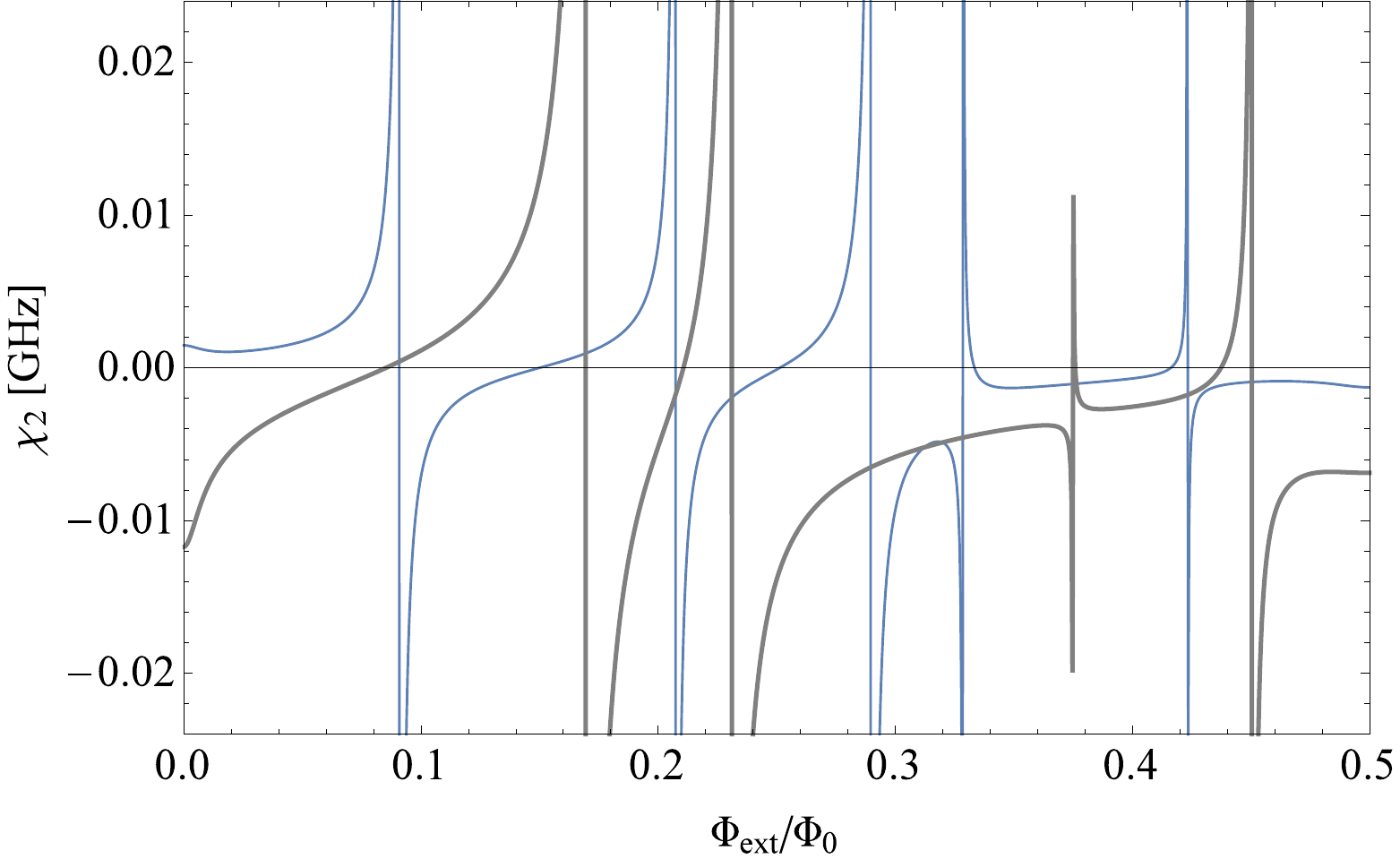}
 \caption{(Color online) Dispersive shift $\chi_2$ [\eref{chir}] as function of external flux for parameter set 1 (thin line) and 2 (thick line).}
 \label{chi2fig}
\end{figure}

The dispersive shifts calculated via \eref{chir} diverge as the energy differences $\omega_{l0}$ or $\omega_{l1}$ approach one of the collective modes frequencies $\omega_\rho$. This divergence, however, only signals the breakdown of the perturbative calculation near such resonant conditions: the actual dispersive shift is limited in magnitude by the coupling constant $g_\rho$, so validity of the perturbative approach is given by the condition $|\chi_\rho| \ll g_\rho$ \cite{flow}. Despite this limitation, \eref{chir} correctly estimates the dispersive shifts at most flux values. We show in Figs.~\ref{chi1fig} and \ref{chi2fig} the flux dependence of dispersive shifts $\chi_1$ and $\chi_2$, respectively. We see that away from resonances the shifts for set 2 are usually larger, as expected due to the larger coupling strengths [cf. Fig.~\ref{grfig}]. Moreover, due to the higher energy of the collective mode involved, the shifts $\chi_2$ displays a richer resonance structure. We will comment on the effect of these shifts on qubit coherence in Sec.~\ref{sec:coher}. Next, we consider the effect of the qubit-mode coupling on the qubit frequency in the absence of excitations of the modes.


\subsection{Qubit frequency renormalization}
\label{sec:qfr}

\begin{figure}[bt]
 \includegraphics[width=0.48\textwidth]{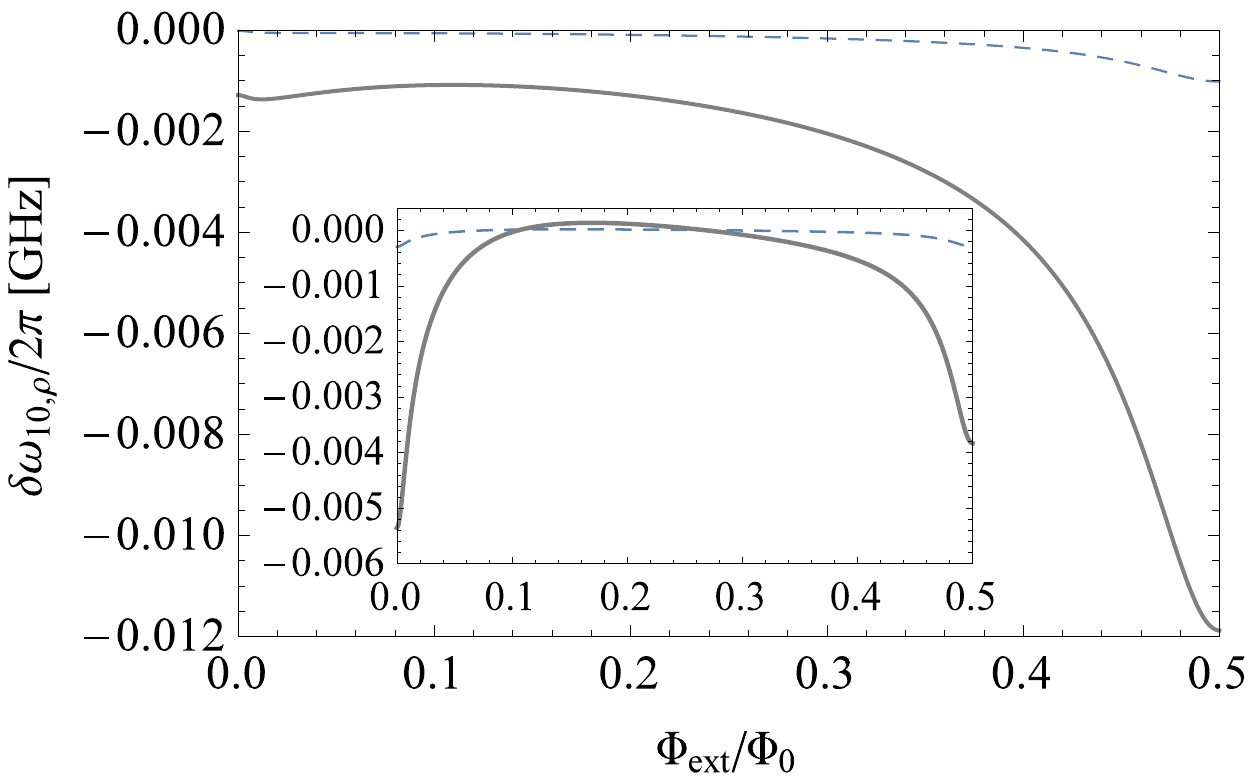}
 \caption{(Color online) Contributions $\delta\omega_{10,1}$ (thick line) and $\delta\omega_{10,2}$ (thin, dashed) [\eref{do10r}] to the qubit frequency renormalization $\delta\omega_{10}$ for parameter set 1 (inset) and 2 (main panel). For comparison, we note that the unrenormalized frequency $\omega_{10}$ approximately varies between 9~GHz (8.2~GHz) at zero flux and 0.33~GHz (0.64~GHz) at half flux quantum for parameter set 1 (set 2).}
 \label{dorfig}
\end{figure}

In the preceding section we studied the change in the qubit frequency when a collective mode is excited, but the interaction term $H_\mathrm{int}$ of \eref{Hint} modifies the qubit frequency even in the absence of collective mode excitations, $\omega_{10} \to \omega_{10} + \delta\omega_{10}$. This frequency correction $\delta\omega_{10}$, arising from Lamb-type energy level shifts, is given by
\be
\delta\omega_{10} = \sum_\rho \delta\omega_{10,\rho}
\ee
with
\be\label{do10r}\begin{split}
\delta\omega_{10,\rho} = \frac12 \sqrt{\frac{E_J^a}{8E^e_{C,\rho}}} \, g_\rho^2 \bigg[ & \left|\langle 0 | p_\phi | 1 \rangle\right|^2 \frac{2\omega_{10}}{\omega_{10}^2 - \omega_\rho^2} \\
+ \sum_{l\ge 2} & \left|\langle 0 | p_\phi | l \rangle\right|^2 \frac{1}{\omega_{l0} + \omega_\rho} \\
- \sum_{l\ge 2} & \left|\langle 1 | p_\phi | l \rangle\right|^2 \frac{1}{\omega_{l1} + \omega_\rho} \bigg].
\end{split}\ee
While this formula resemble \eref{chir}, there is one important difference: so long as $\omega_\rho > \omega_{10}$, there are no divergences in \eref{do10r}. In Fig.~\ref{dorfig} we plot the first two largest contributions to $\delta\omega_{10}$, namely $\delta\omega_{10,1}$ and $\delta\omega_{10,2}$. The former is generally much larger (in absolute value) than the latter, due to the stronger coupling [cf. Fig.~\ref{grfig}]. Note that even at half flux quantum, where $\omega_{10}$ has a minimum \cite{flux2,nature}, the correction is at most a few percent of $\omega_{10}$.

\subsection{Purcell rate}
\label{sec:purc}

So far we have considered the system to be capacitively coupled to external voltage sources. In practical realizations of circuit QED experiments, this coupling is to a mode of a cavity; this can be accounted for by replacing \cite{pratr}
\be
V \to \tilde{V} (c^\dagger + c)
\ee
in the coupling Hamiltonian $H_V$, \eref{HV}. Here parameter $\tilde{V}$ accounts for the strength of the electric field at the qubit position as well as for the geometry of the cavity-qubit system, while $c^\dagger$ ($c$) are creation (annihilation) operators for photons in the cavity. As it is customary, to include the cavity and its coupling to an external bath of harmonic oscillators, we add to $H_e$ in \eref{He} the following Hamiltonian $H_{cb}$
\be
H_{cb} = \omega_c c^\dagger c + \sum_k \omega_k b^\dagger_k b_k + \sum_k \lambda_k \left(b^\dagger_k c + c^\dagger b_k \right),
\ee
where $\omega_c$ is the cavity frequency, $b^\dagger_k$ ($b_k$) are creation (annihilation) operators for bath excitations with energy $\omega_k$, and $\lambda_k$ the coupling strengths between cavity and bath modes.

\begin{figure}[bt]
 \includegraphics[width=0.48\textwidth]{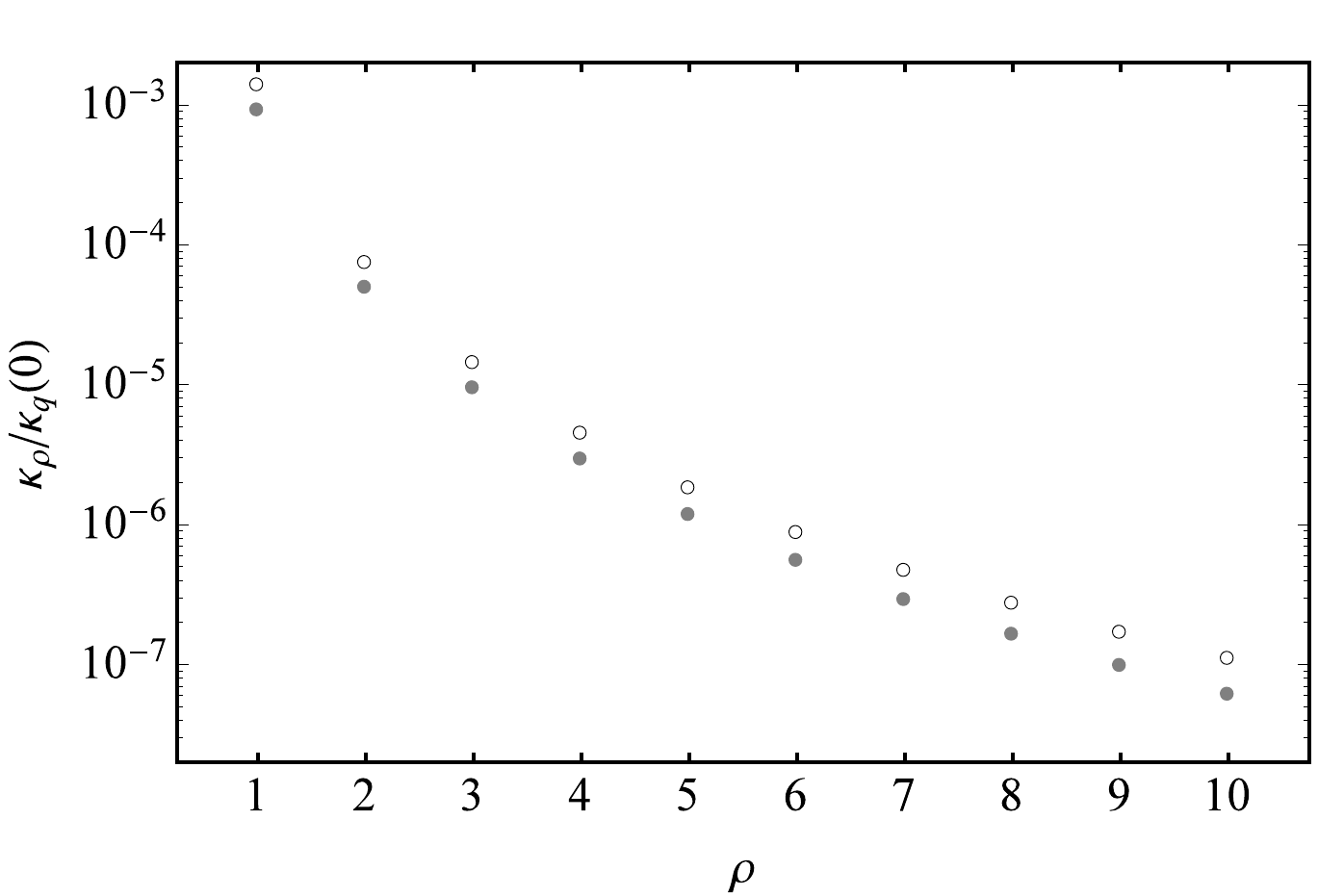}
 \caption{Ratio $\kappa_\rho/\kappa_q(0)$ between collective mode and qubit Purcell decay rates [\esref{Grho} and \rref{Gq}, respectively] for the ten lowest even modes. Filled circles: parameter set 1 in Table~\ref{tabpar}; empty circles: set 2.}
 \label{grq}
\end{figure}

Within this model, one can calculate using Fermi's golden ruled the so-called Purcell rate $\kappa_q$ for the qubit; that is, the decay rate of the qubit excited state by emission of a photon into the bath (mediated by the cavity):
\be\label{Gq}
\kappa_q(f) = \kappa \frac{g_q^2(f)}{\left(\omega_{10}(f) - \omega_c \right)^2} \, ,
\ee
where $\kappa$ is the inverse lifetime of a photon in the cavity, as determined by the cavity-bath couplings $\lambda_k$ \cite{pratr}, and
\be
g_q= 2e\tilde{V} \frac{\tilde{E}_C^\phi}{E_C^c} \langle 0 | p_\phi | 1 \rangle \, .
\ee
This coupling constant depends on flux via the qubit states.
The above expression for $\kappa_q$ is valid in the dispersive regime $g_q \ll |\omega_{10} - \omega_c|$.
The similar calculation for the collective modes gives their decay rate as
\be\label{Grho}
\kappa_\rho = \kappa \frac{1}{\left(\omega_\rho - \omega_c \right)^2} \left[\frac{e\tilde{V}}{4E_C^c} g_\rho \frac{1}{\sqrt{2} \ell_\rho} \right]^2,
\ee
where $\ell_\rho=(8E_{C,\rho}^e/E_J^a)^{1/4}$ and again we have assumed that the factor multiplying $\kappa$ is small compared to unity. Note that the term in square brackets is a decreasing function of $\rho$; therefore, so long as all the modes have frequency above the cavity one ($\omega_1 > \omega_c$), the decay rate of the collective modes decreases with $\rho$. If the qubit lifetime is limited by Purcell relaxation, we can then estimate the modes' lifetimes by eliminating the unknown quantity $\tilde{V}$ from \esref{Gq} and \rref{Grho}. In fact, the ratio
ratio $\kappa_\rho/\kappa_q(f)$ is also independent of $\kappa$ and is determined by the circuit properties and the cavity frequency; we plot $\kappa_\rho/\kappa_q(0)$ in Fig.~\ref{grq} for the parameters given in Table~\ref{tabpar}, with $\omega_c/2\pi=8.18$~GHz for set 1 and $\omega_c/2\pi=8.89$~GHz for set 2 [see Refs.~\onlinecite{flux1} and \onlinecite{nature}, respectively].

\section{Non-linearity of the array junctions}
\label{sec:anhar}

By expanding the last term in \eref{US} to quadratic order to obtain the approximate potential energy term in \eref{U}, we have treated the Josephson junctions in the array as
linear elements. However, the cosine in \eref{US} includes their non-linear properties, and in this section we account perturbatively for these non-linearities. To begin with, we split the Josephson energy of each junction into two contributions using the identity
\be\label{cid}\begin{split}
\cos\left[\frac{\phi}{N}+\sum_\mu W_{\mu m}\xi_\mu \right] = \cos\frac{\phi}{N} \cos \left[ \sum_\mu W_{\mu m}\xi_\mu \right]\phantom{.} \\
- \sin\frac{\phi}{N} \sin \left[ \sum_\mu W_{\mu m}\xi_\mu \right].
\end{split}\ee
The sine product term, as we discuss below, generates two qubit-collective mode interaction terms that we will denote with $U^{(1)}_{\phi\xi}$ and $U^{(3)}_{\phi\xi}$.
The product of the two cosines can be rewritten as:
\be\label{cceq}\begin{split}
&\cos\frac{\phi}{N} \cos \left[ \sum_\mu W_{\mu m}\xi_\mu \right] = \cos\frac{\phi}{N} \bar{\cal C}_{0m} \\
&+  \cos \left[ \sum_\mu W_{\mu m}\xi_\mu \right] -  \bar{\cal C}_{0m} \\
&+ \left(\cos\frac{\phi}{N} - 1 \right) \left( \cos \left[ \sum_\mu W_{\mu m}\xi_\mu \right] -  \bar{\cal C}_{0m} \right),
\end{split}\ee
where
\be\label{C0}
\bar{\cal C}_{0m} = \left\langle\cos \left[ \sum_\mu W_{\mu m}\xi_\mu \right]\right\rangle_0
\ee
is the expectation value of the operator inside the angular brackets in the ground state of the collective modes. As detailed in the next section, the first term in the right hand side of \eref{cceq} gives rise to the qubit mode effective potential $U_\phi$, while the last term is a qubit-collective modes interaction contribution, denoted with $U^{(2)}_{\phi\xi}$. The third term is a constant that can be neglected. The second term gives, to lowest order, the harmonic potential energy of the collective modes,
\be
\sum_m \cos \left[ \sum_\mu W_{\mu m}\xi_\mu \right] \approx N -\frac12 \sum_\mu \xi_\mu^2 + \ldots
\ee
The higher order terms in the expansion neglected here leads to interactions among the collective modes that do not affect the qubit directly. We do not consider such interactions from now on, and write the potential energy $U_S$ in the approximate form
\be
U_S \simeq  \frac12 E_J^a \sum_\mu \xi_\mu^2 + U_\phi + \sum_{j=1}^3 U^{(j)}_{\phi\xi} \,
\ee
with the potentials $U_{\phi}$ and $U^{(j)}_{\phi\xi}$ specified in what follows.

\subsection{Qubit effective potential $U_\phi$}
\label{sec:effpot}

Keeping only the first term in the right hand side of \eref{cceq}, from \eref{US} we find
\be
U_\phi = -E_J^b \cos \left(\phi+\varphi_e\right) - E_L \left[N^2 \cos \frac{\phi}{N} \right] \bar{\cal C}_{0}
\ee
with $\bar{\cal C}_0 = \sum_m \bar{\cal C}_{0m}/N$. Upon expansion of the term in square bracket (valid for $|\phi| \ll \pi N$) and assuming $\bar{\cal C}_{0} \approx 1$, we recover the quadratic inductive energy term $E_L \phi^2/2$ of \eref{U}. The full qubit potential, however, contains small additional non-linearities originating from the higher-order terms of the expansion. Here we do not consider these terms further, as they are suppressed by factors of the form $1/N^{2j}$ ($j=1,\,2,\, 3, \ldots$), but we show that in general $\bar{\cal C}_{0} < 1$; therefore, the actual inductive energy
\be\label{tEl}
\tilde{E}_L = E_L \bar{\cal C}_0
\ee
is smaller than what the simple expression $E_L=E_J^a/N$ suggests.

The expectation value entering $\bar{\cal C}_{0m}$, \eref{C0}, can be readily obtained from the known matrix elements for the harmonic oscillator, see for example Appendix~D of \ocite{prb1}. Here we have to remember that in the odd sector a rotation from the original modes $\zeta_\rho$ to
independent modes $\tilde\zeta_\rho$ is necessary, see Sec.~\ref{sec:odd}; this is accomplished via the orthogonal matrix $\Lambda_{\mu\nu}$ defined in \eref{LMdef}. We thus arrive at
\be
\bar{\cal C}_{0} =\frac{1}{N} \sum_m \exp\left[-\frac14 \sum_{\mu,\nu,\tau} W_{\mu m} W_{\nu m} \Lambda_{\tau\mu}
\Lambda_{\tau\nu} \ell^2_\tau \right],
\ee
where
\be
\ell_\tau = \left(8E_{C,\tau}/E_J^a\right)^{1/4}
\ee
is the oscillator length for mode $\tau$, with $E_{C,\tau}$ given in \esref{ecodd} and \rref{eceven} for odd and even modes, respectively. Clearly $\bar{\cal C}_0 <1$ so long as at least one oscillator length is finite. We can also find a lower bound (and rough estimate) for $\bar{\cal C}_0$ by noting that $\ell_\tau \le \ell_0$, where $\ell_0 = \left(8E_{C}^a/E_J^a\right)^{1/4}$ is the oscillator length in the absence of capacitance to ground in the array; note that we typically have $\ell_0 \lesssim 1$, cf. Table~\ref{tabpar}. Then, using the identities $\sum_\tau \Lambda^2_{\tau\nu}=1$ and $\sum_\mu W_{\mu m}^2 = (N-1)/N$, we find:
\be
\bar{\cal C}_0 \ge \exp \left[-\ell_0^2 \frac{N-1}{4N}\right].
\ee
The expansion to lowest order in $\ell_0^2$ of this formula agrees with the expression for the reduction of $E_L$ reported in \ocite{prx}. Our result shows that that expression generally overestimates the suppression of the inductive energy.

\subsection{Quadratic interaction $U^{(2)}_{\phi\xi}$}
\label{sec:u2}

We now consider the leading order contribution to the potential energy originating from the last term in \eref{cceq}. By expanding the term dependent on the collective mode coordinates and introducing the creation/annhilation operators via $\xi_\mu = \ell_\mu (a_\mu + a^\dagger_\mu)/\sqrt{2}$ we arrive at
\be\label{U2}
U^{(2)}_{\phi\xi} = \frac{E_J^a}{2}\left(1-\cos\frac{\phi}{N} \right) \sum_\mu \frac{\ell_\mu^2}{2} \left( 2 a^\dagger_\mu a_\mu +  a^\dagger_\mu a^\dagger_\mu + a_\mu a_\mu \right).
\ee
Note that in the absence of the array ground capacitances (so that $\ell_\mu \equiv \ell_0$) this interaction term is invariant under orthogonal transformations belonging to the group $\mathrm{O}(N-1)$ -- the $\mathrm{U}(N-1)$ symmetry of approximate Lagrangian ${\cal L}_U$ in \eref{LU} is only partially broken (the first term in brackets actually fully preserves $\mathrm{U}(N-1)$ symmetry).

\begin{figure}[bt]
 \includegraphics[width=0.48\textwidth]{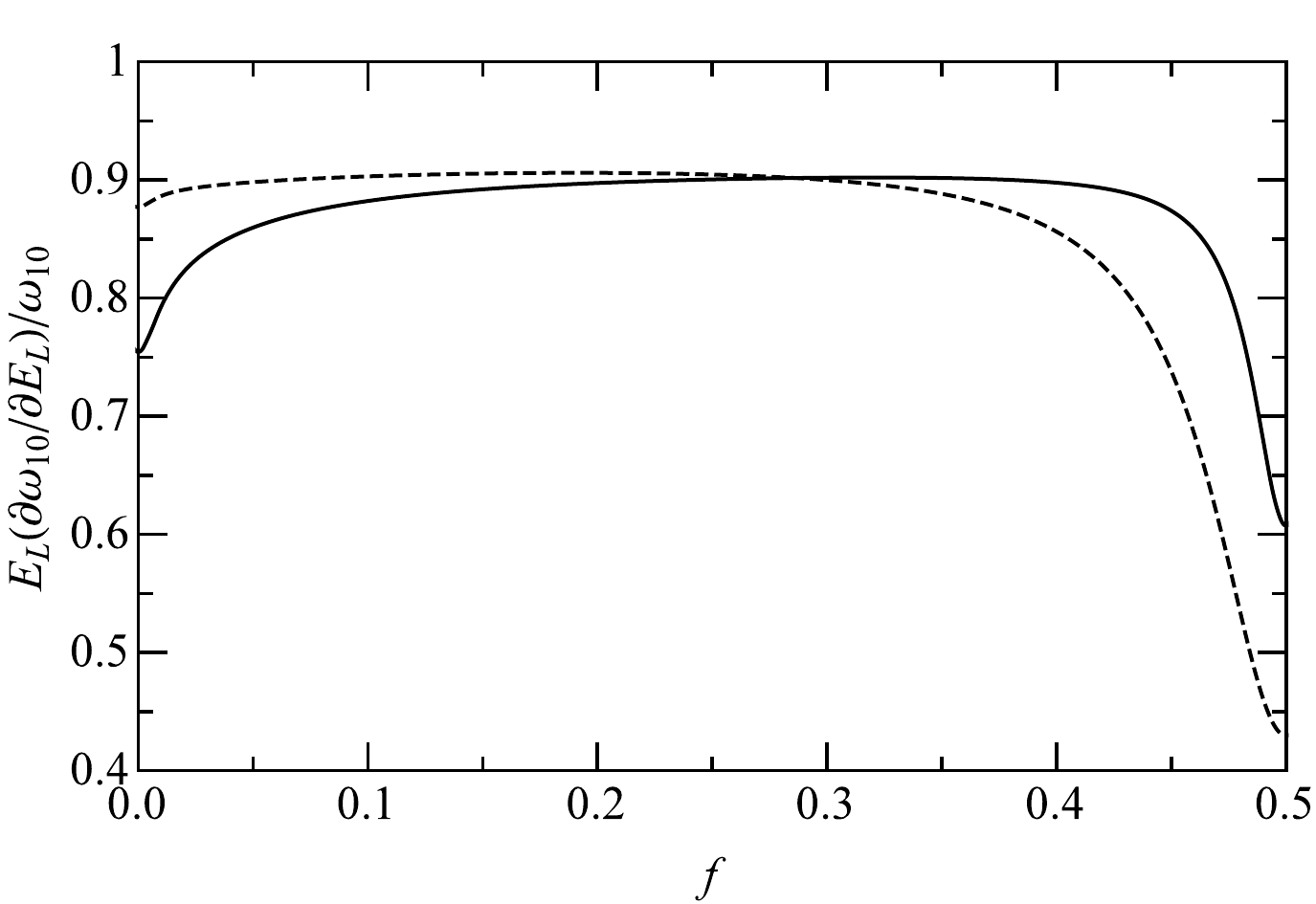}
 \caption{Normalized derivative $\partial \omega_{10}/\partial E_L$ calculated for the parameters in Table~\ref{tabpar}. Solid line: set 1; dashed line: set 2. Note the minima at $f=0$ and $f=1/2$.}
 \label{dodel}
\end{figure}

In \eref{U2} we can distinguish two contributions. First, there are terms which are proportional to each collective modes number operator $n_\mu = a^\dagger_\mu a_\mu$; to lowest order in $1/N$ these terms are
\be
U_{\delta E_L} = \frac12 \phi^2 \left[ \frac{E_L}{2N} \sum_\mu \ell_\mu^2 a^\dagger_\mu a_\mu \right],
\ee
and they give a dependence of the inductive energy $E_L$ on the occupation of the collective modes. Since changes in $E_L$ lead to variations of the qubit frequency, this dependence can be interpreted as a dispersive shift $\chi^{\delta E_L}_\mu$:
\be\label{chidel}
\chi^{\delta E_L}_{\mu} = \frac12 \left(\frac{\partial \omega_{10}}{\partial E_L}\right) \frac{E_L}{2N} \ell^2_\mu\, .
\ee
Over a broad range of fluxes, except near half-integer multiples of the flux quantum, the qubit frequency is approximately proportional to the inductive energy \cite{prb1},
\be\label{o10el}
\omega_{10} \approx (2\pi)^2 E_L \left| f- \frac12\right| \, ,
\ee
so that $E_L \left(\partial \omega_{10}/\partial E_L\right) \approx \omega_{10}$. This approximate relation translates at all fluxes in an upper bound for the dispersive shifts:
\be\label{chidelb}
\chi^{\delta E_L}_{\mu} \lesssim \frac{\omega_{10}}{4N} \ell^2_\mu \, .
\ee
This bound shows that the dispersive shifts lead to relative changes of order $1/N$, in the qubit frequency. Interestingly, the derivative $\partial \omega_{10}/\partial E_L$ and hence the dispersive shifts have minima at half-integer multiples of the flux quantum, see Fig.~\ref{dodel}.
Therefore the dephasing induced by $U_{\delta E_L}$ is suppressed at these ``sweet spots'', similar to the suppression of dephasing by flux noise; this is not surprising, since at leading order the flux and inductive energy affect the qubit frequency in the same way, see \eref{o10el}. However, for typical experimental parameters
the reduction is less than one order of magnitude.

The second type of contribution in $U^{(2)}_{\phi\xi}$ comes from terms of the form $a_\mu a_\mu + a^\dagger_\mu a^\dagger_\mu$. As in Sec.~\ref{sec:even}, the effect of such terms can be studied by performing a Schrieffer-Wolff transformation, as detailed in Appendix~\ref{app:SWU2}. Here we simply note that, since they involve the virtual exchange of two collective mode excitations rather than one, the resulting dispersive shifts are generally smaller than $\chi_\mu^{\delta E_L}$ of \eref{chidel} and can therefore be neglected.

\subsection{Linear interaction $U^{(1)}_{\phi\xi}$}

In this subsection and the next one, we focus on the perturbative treatment of the last term in \eref{cid}, obtained by expanding the sine with argument the collective modes coordinates. The contribution to the potential energy from the linear term in this expansion vanishes by construction, due to the property $\sum_m W_{\mu m} =0$. As we show in Appendix~\ref{app:U1U3}, the third order term gives rise to two types of interactions, one of them being a linear interaction term of the form
\be\label{U1}\begin{split}
U^{(1)}_{\phi\xi} & = \sum_{\rho=1}^{N_e} \tilde{g}_\rho \left(N \sin \frac{\phi}{N}\right) \eta_\rho \, , \\
\tilde{g}_\rho & = \frac{1}{2(2N)^{3/2}} E_J^a \left[\ell^2_{N-\rho} - \ell^2_{\rho}\right]
\end{split}\ee
with $\eta_\rho$ and $N_e$ defined in the text after \eref{Leven}.
Since it couples the qubit mode with even collective modes only, this interaction preserves $PT$-symmetry. Moreover, in the $\mathrm{S}_N$ symmetric case (i.e., neglecting ground capacitances, so that $\ell_\mu \equiv \ell_0$ for all $\mu$) this term is absent, in agreement with the group-theoretical analysis of \ocite{prx}.
The coupling constant $\tilde{g}_\rho$ decreases with the collective mode index $\rho$, albeit more slowly than $g_\rho$ in \eref{Hint} for low $\rho$, while for large index $\rho \lesssim N_e$ we find
\be\label{gtgra}
\frac{\tilde{g}_\rho}{g_\rho} \approx \frac{\omega_p^a}{32 N \tilde{E}_C^\phi} \, ,
\ee
where $\omega_p^a = \sqrt{8E_J^a E_C^a}$ is the array junction plasma frequency. A (loose) upper bound for $\tilde{g}_1$ is given by
\be
\tilde{g}_1 < \frac{E_J^a}{2} \frac{\ell^2_{0}}{(2N)^{3/2}} = \frac{1}{2} \frac{\omega_p^a}{(2N)^{3/2}} \, .
\ee

\begin{figure}[bt]
 \includegraphics[width=0.48\textwidth]{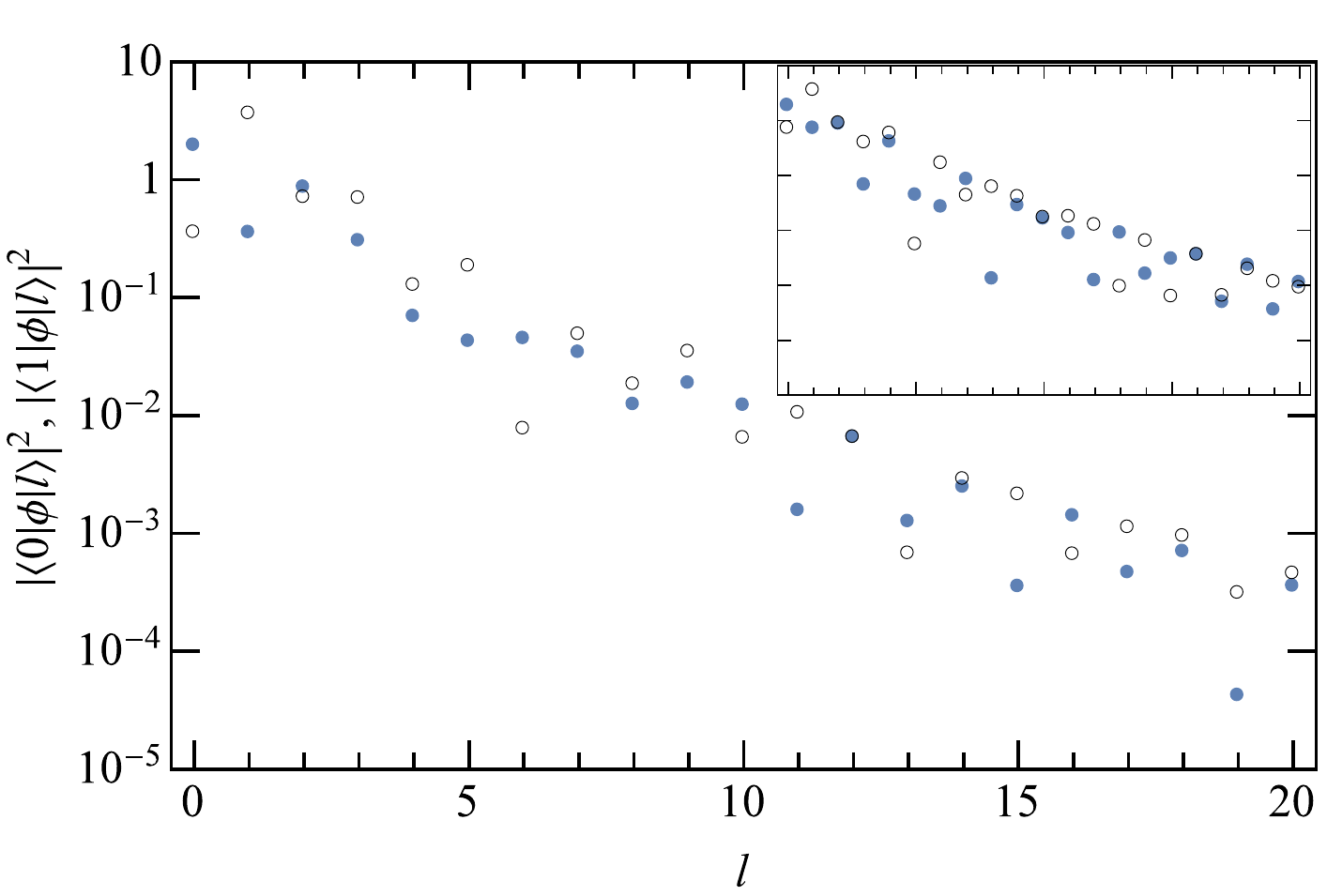}
 \caption{(Color online) Main panel: matrix elements squared $|\langle 0 | \phi | l \rangle|^2$ (filled circles) and  $|\langle 1 | \phi | l \rangle|^2$ (empty circles) at $f=0.35$ for $l\le 20$ calculated using parameter set 1 [see Table~\ref{tabpar}]. Inset: same as main panel but for parameter set 2.}
 \label{xme}
\end{figure}

\begin{figure}[tb]
 \includegraphics[width=0.48\textwidth]{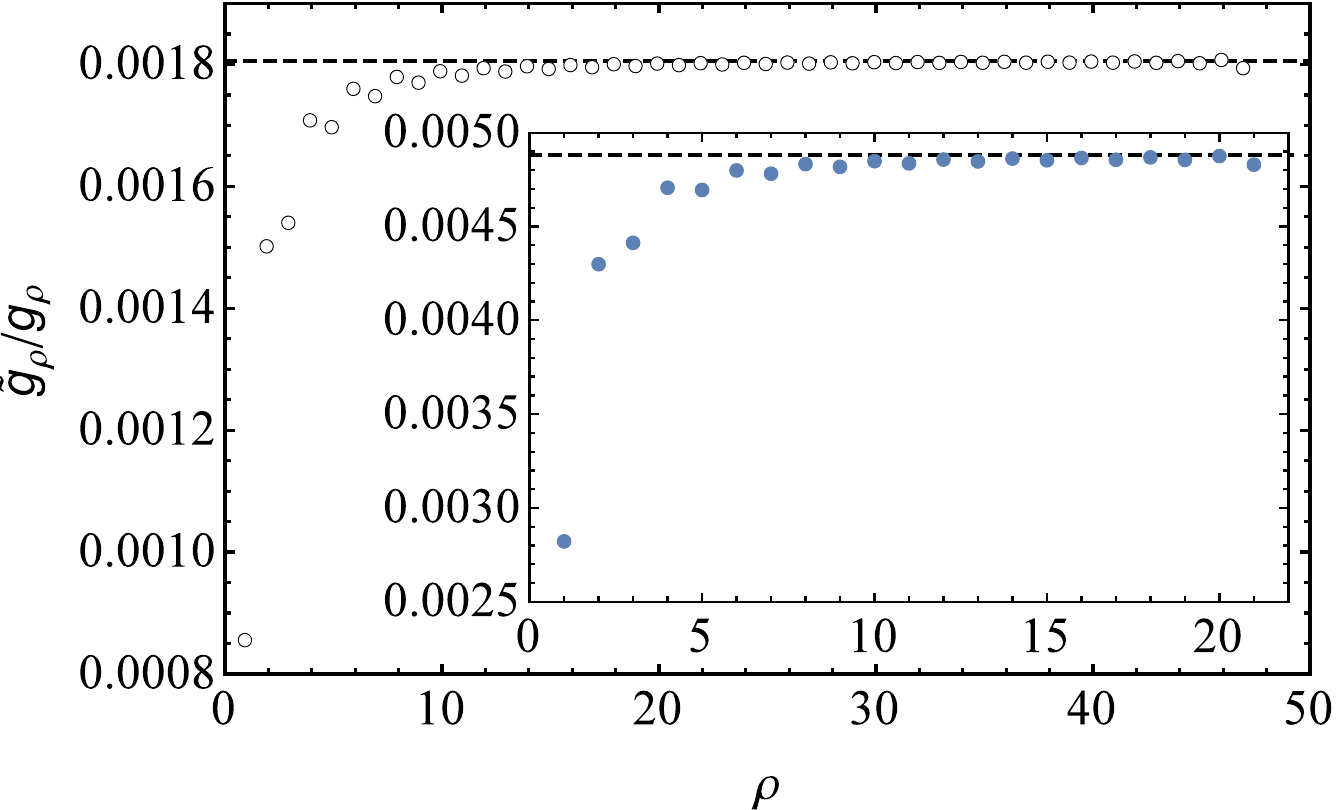}
 \caption{Coupling constants ratio $\tilde{g}_\rho/g_\rho$ for parameter set 1 (inset, filled circles) and 2 (main panel, empty circles) for all the even modes. Horizontal dashed lines are given by the right hand side of \eref{gtgra}.}
 \label{gtgr}
\end{figure}

As done for the interaction term $H_\mathrm{int}$ in \eref{Hint}, the effect of $U^{(1)}_{\phi\xi}$ on the qubit can be more easily studied by performing a Schrieffer-Wolff transformation leading to the additional dispersive shift $\chi_\rho^{(1)}(f)$:
\be\label{chir1}\begin{split}
\chi_\rho^{(1)} = \frac12 \sqrt{\frac{8E^e_{C,\rho}}{E_J^a}} \,
\tilde{g}_\rho^2 \bigg[ & \left|\langle 0 | \left(N \sin \frac{\phi}{N}\right) | 1 \rangle\right|^2 \frac{2\omega_{10}}{\omega_{10}^2 - \omega_\rho^2} \\
+ \sum_{l\ge 2} & \left|\langle 0 | \left(N \sin \frac{\phi}{N}\right) | l \rangle\right|^2 \frac{\omega_{l0}}{\omega_{l0}^2 - \omega_\rho^2} \\
- \sum_{l\ge 2} & \left|\langle 1 | \left(N \sin \frac{\phi}{N}\right) | l \rangle\right|^2 \frac{\omega_{l1}}{\omega_{l1}^2 - \omega_\rho^2} \bigg].
\end{split}\ee
A few comments are in order: first, since the low-lying states are localized (so that relevant values of $\phi$ are at most of order $2\pi$) and  $N$ is large, a good approximation for the matrix elements is obtained with the substitution $N \sin \phi/N \to \phi$. Second,
within this approximation, numerical calculation of $\chi_\rho^{(1)}$ does not require much additional computation compared to that of $\chi_\rho$, \eref{chir}, thanks to the identity \cite{fid}
\be\label{meid}
\omega_{ml}^2\left|\langle m | \phi | l \rangle\right|^2 = \left(8\tilde{E}_C^\phi\right)^2 \left|\langle m | p_\phi | l \rangle\right|^2\, .
\ee
Third, the pole structure in \eref{chir1} is the same as that of $\chi_\rho$ and the matrix elements again becomes smaller as $l$ increases, see Fig.~\ref{xme} and \eref{meid}. Finally, for typical experimental parameters, the coupling constants $\tilde{g}_\rho$ are at least two orders of magnitude smaller than $g_\rho$ [see Fig.~\ref{gtgr}], so we expect the dispersive shifts $\chi^{(1)}_\rho$ to have negligible effect -- we will return to this point in Sec.~\ref{sec:coher}.

\subsection{Multi-mode interaction $U^{(3)}_{\phi\xi}$}

All the interactions discussed so far involve the qubit mode and a single collective mode. This is not the case for the interaction term $U^{(3)}_{\phi\xi}$, which involves up to three collective modes:
\be\label{U3}\begin{split}
& U^{(3)}_{\phi\xi} = -\frac{E_J^a}{24\sqrt{N}} \sin\frac{\phi}{N} \bigg[ 3 \sum_{\mu,\nu=1}^{\mu+\nu < N}  \ell_\mu \ell_\nu \ell_{\mu+\nu} \Big( a^\dagger_{\mu+\nu} a^\dagger_\nu a^\dagger_\mu \\ & + 2 a^\dagger_{\mu+\nu} a^\dagger_\nu a_\mu + a^\dagger_{\nu} a^\dagger_\mu a_{\mu+\nu} + \mathrm{H.c.} \Big) - \sum_{\mu+\nu > N}^{N-1} \ell_\mu \ell_\nu \\ &\times \ell_{2N-\mu-\nu} \Big(a^\dagger_{2N - \mu-\nu} a^\dagger_\nu a^\dagger_\mu +
3 a^\dagger_{2N - \mu-\nu} a^\dagger_\nu a_\mu + \mathrm{H.c.}\Big)\bigg],
\end{split}\ee
where H.c. denotes the Hermitian conjugate. The first term, for example, contains creation operators of two modes if $\mu = \nu$ and three modes if $\mu \neq \nu$. Note that the index structure ensures that at least one index is even and the remaining two indices have the same parity; this shows that $PT$-symmetry is preserved.

As a consequence of the presence of three creation-annihilation operators in \eref{U3}, a description in terms of an effective Hamiltonian would involve terms with products of up to three number operators (see also the discussion of the $U^{(2)}_{\phi\xi}$ interaction in Appendix~\ref{app:SWU2}). Rather than attempting such a complicated description here, we consider the case in which the occupation probability of each mode is sufficiently small that we can neglect the possibility of having two or more excitations in a mode or two or more modes being excited at the same time; this requires the occupation probability to be small compared to $1/N$, see Appendix~\ref{app:occ}. In other words, we only consider the possibility that no more than one collective mode is excited at any given time.
We can then calculate the change $\Delta\omega_{10,\mu}$ in qubit frequency from when the collective modes are in their ground state $|0\rangle$ ($a_\mu|0\rangle = 0$ for any $\mu$) to when one of the collective modes is excited, i.e., in state $|1_\mu\rangle = a^\dagger_\mu | 0 \rangle$. Such a frequency change resembles the dispersive shifts discussed so far, although those are valid for multiple excitations in each mode. The perturbative calculation of the frequency change is detailed in Appendix~\ref{app:U1U3};
there we also show that the frequency change is smaller than the dispersive shift $\chi_\mu^{\delta E_L}$ in \eref{chidel} and can therefore be neglected.
In the next section we explore some effects of the collective modes dispersive shifts on the qubit.

\section{Qubit dephasing}
\label{sec:coher}

\begin{figure}[tb]
 \includegraphics[width=0.48\textwidth]{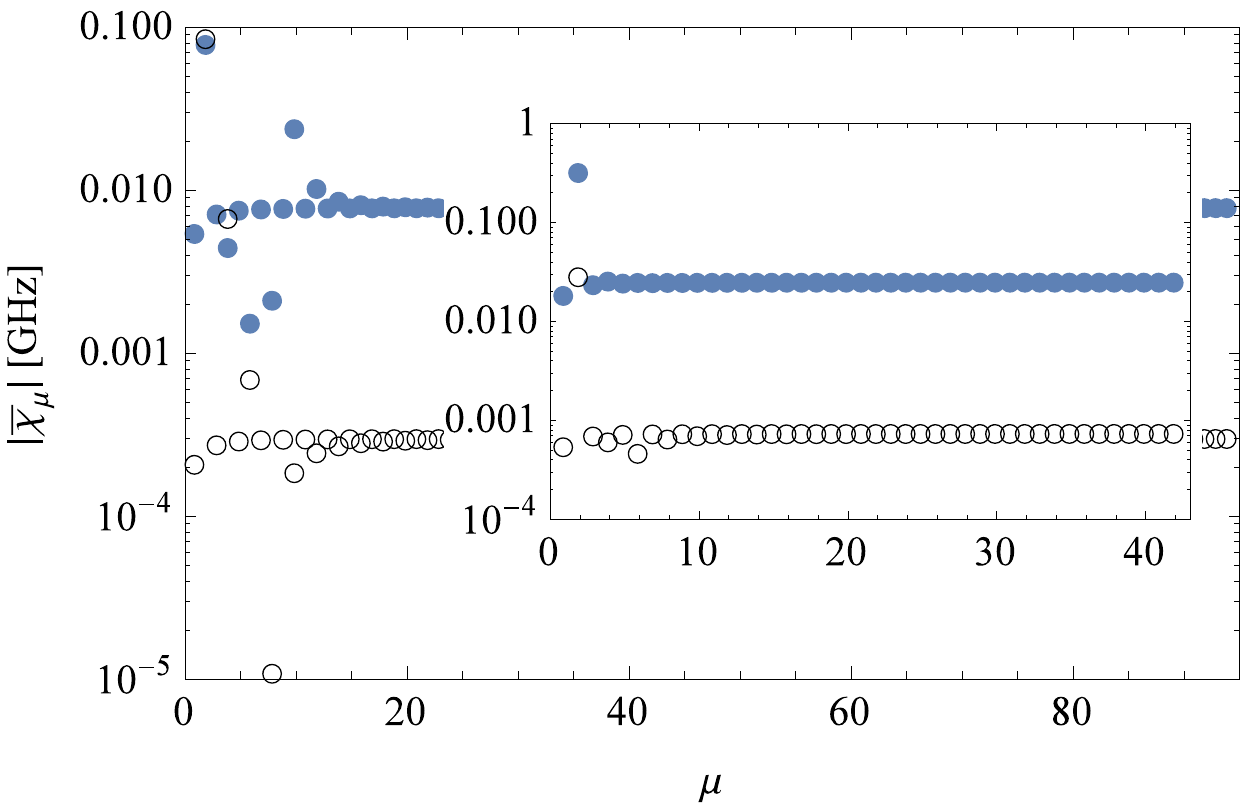}
 \caption{Absolute values of the total dispersive shifts $|\bar\chi_\mu|$ for parameter set 1 (inset) and 2 (main panel). Filled circles are evaluated at zero external flux, empty circles at half flux quantum. The higher collective modes have very similar values of $|\bar\chi_\mu|$, dominated by $\chi_\mu^{\delta E_L}$ of \eref{chidel}. The large fluctuations of $|\bar\chi_\mu|$ for the low even modes are due to $\chi_\rho$ of \eref{chir}. In all cases, the lowest even mode ($\mu = 2$) has the largest shift.}
 \label{chibar}
\end{figure}

As it is well known, the last term in the effective Hamiltonian \eref{Heff} can be interpreted as a shift in the qubit frequency dependent on the states of the collective modes; more generally, we write
\be
\omega_{10}\left(\{n_\mu\}\right) = \omega_{10} + 2\sum_\mu \chi_\mu^t n_\mu \, ,
\ee
where $n_\mu$ is the occupation number of mode $\mu$ and we use $\bar\chi_\mu$ to denote the total dispersive shift of that mode. For even $\mu=2\rho$ it is given by
\be
\bar{\chi}_{2\rho} = \chi_{\rho} + \chi_{\rho}^{(1)} + \chi_{2\rho}^{\delta E_L}
\ee
while $\bar\chi_\mu = \chi_\mu^{\delta E_L}$ for $\mu$ odd. Therefore for even modes the total shift includes the contributions from \eref{chir} and \rref{chir1} together with that in \eref{chidel}. However, we note that for almost all values of flux \cite{notechiflux} we have $|\chi_\rho^{(1)}/\chi_\rho|\sim (\tilde{g}_\rho/g_\rho)^2$ and the latter quantity is $\lesssim 10^{-5}$ for experimentally relevant parameters, see Fig.~\ref{gtgr}; therefore, we neglect $\chi_\rho^{(1)}$ from now on. As for relative importance of the contributions $\chi_\rho$ and $\chi_{2\rho}^{\delta E_L}$, we note that while the former quickly decreases in magnitude with increasing index $\rho$, the latter slowly increases with $\rho$. Therefore we can expect that while it may be necessary to keep $\chi_\rho$ for the low index modes, $\chi_{2\rho}^{\delta E_L}$ could be the only relevant contribution for higher index modes. To identify which modes have low index in this sense, we remind that in the dispersive regime $|\chi_\rho| \ll g_\rho$, so we can certainly neglect $\chi_\rho$ if $g_\rho < |\chi_{2\rho}^{\delta E_L}|$. Unfortunately this latter condition is satisfied at any flux only for high index modes,
so in general we must keep both $\chi_\rho$ and $\chi_{2\rho}^{\delta E_L}$ for quantitative estimates -- see also Fig.~\ref{chibar}.

The fluctuations of the occupations of the collective modes cause fluctuations in the qubit frequency and hence dephasing. For the qubit-cavity coupling, the so-called photon shot noise dephasing has been investigated in the 3D transmon architecture \cite{psn,rig}. In particular, in \ocite{psn} good agreement between theory and experiment was found over a range of average occupation number in the cavity $\bar{n}$ from small to relatively large ($\bar{n} \sim 3$), and a residual occupation of order 1\% was estimated. Since the collective modes are (weakly) coupled to the cavity and the residual occupation is small, here we restrict ourselves to the relevant case of small occupation number $\bar{n}_\mu \ll 1$ for any collective mode. In this case
each mode contributes a rate \cite{rig}
\be\label{Garho}
\Gamma_\mu = \frac{4\kappa_\mu \bar\chi_\mu^2}{\kappa_\mu^2+ 4\bar\chi_\mu^2} \bar{n}_\mu \, ,
\ee
with $\kappa_\mu$ the decay rate of mode $\mu$ \cite{fdeph}, to the total qubit dephasing rate
\be\label{Gaphi}
\Gamma_\phi = \sum_\mu \Gamma_\mu \, .
\ee
Equation \rref{Garho} was derived assuming that the effect of each mode can be treated independently \cite{rig}.
This expression enables us to put an upper limit on the dephasing rate, since independently of $\kappa_\mu$, the rate satisfies $\Gamma_\mu \le |\bar\chi_\mu| \bar{n}_\mu$. The inequality is saturated for $\kappa_\mu = 2|\bar\chi_\mu|$, and becomes a strong upper bound in the limiting cases of $\kappa_\mu$ much bigger or smaller than $|\bar\chi_\mu|$.
In the remaining of this section we restrict our attention to flux being zero or half a flux quantum, since it is experimentally established \cite{flux2} that away from these ``sweet spots'' the fluxonium dephasing rate is determined by flux noise. However, we note that near resonances where the dispersive shifts are enhanced, see Figs.~\ref{chi1fig} and \ref{chi2fig}, they could give rise to reproducible suppressions of coherence time $T_2$ at specific flux values.

To determine if the collective modes can at least in principle be a significant source of dephasing, let us consider a worst-case scenario in which each rate $\Gamma_\rho$ attains its maximum value and all collective modes are equally populated, $\bar{n}_\mu \equiv \bar{n}$. Then we have
\be\label{gpw}
\Gamma_\phi^w (f) =\bar{n} \sum_\mu \left|\bar\chi_\mu(f) \right| \, .
\ee
We have calculated $|\bar\chi_\mu|$ at zero and half flux quantum for both parameter sets in Table~\ref{tabpar}, see Fig.~\ref{chibar}. Summing over all modes and assuming $\bar{n} = 0.01$, we arrive at the results summarized in Table~\ref{tabdeph}. We find that in the worst case, the collective modes could limit the dephasing time  to about $0.1 \mu$s at zero flux and $1\mu$s at half flux quantum. Measured coherence times are longer than these estimates \cite{flux2,nature}, indicating that the worst case is not realized in practice. In fact, in \ocite{nature} the Purcell-limited lifetime of the qubit at zero flux was measured to be at least $10\mu$s; then the results of Sec.~\ref{sec:purc} and Fig.~\ref{grq} indicate that the decay rate of the lowest even mode is of order 100~Hz, much smaller than the dispersive shift, and all other modes have even smaller decay rates \cite{fodd}. In the more realistic limit $\kappa_\mu \ll |\bar\chi_\mu|$, from \esref{Garho}-\rref{Gaphi} we find
\be\label{gpk}
\Gamma_\phi^\kappa =\bar{n} \sum_\mu \kappa_\mu \, ,
\ee
from which we get the estimates in the last column of Table~\ref{tabdeph}, corresponding to a dephasing time of order 1~s (dominated by the relaxation rate of the lowest even collective mode). This time scale is much longer than the coherence times measured in experiments, indicating that most likely the collective modes are not causing any significant dephasing.
We caution the reader that the estimates in the last column of Table~\ref{tabdeph} rest mainly on the assumption (valid within our model for typical parameter values as in Table~\ref{tabpar}) that the collective modes are much more weakly coupled to the cavity than the qubit is, as in Fig.~\ref{grq}; if the assumption is not correct, this could result in a dephasing time shorter by several orders of magnitudes -- see also the end of Sec.~\ref{sec:sic}.

\begin{table}[bt]
\begin{tabular}{|c||c|c||c|}
\hline
& $\Gamma_\phi^w (0)$ & $\Gamma_\phi^w (0.5)$ & $\Gamma_\phi^\kappa$ \\ \hline
set 1 & 13.1 & 0.56 & $1 \times 10^{-6}$ \\
set 2 & 7.99 & 1.17 & $1.5 \times 10^{-6}$ \\
\hline
\end{tabular}
\caption{Estimates (in MHz) for the dephasing rates $\Gamma_\phi^w$, \eref{gpw}, and $\Gamma_\phi^\kappa$, \eref{gpk} -- see text for details.}
\label{tabdeph}
\end{table}

\section{Broken $PT$-symmetry: an example}
\label{sec:sb}

In all the previous sections we have assumed the system to be $PT$-symmetric, which ensures the decoupling of qubit and odd collective modes in the (approximate) quadratic Lagrangian. In practice it is difficult to fabricate a perfectly symmetric circuit, so it is interesting to investigate what are the main qualitative consequences of breaking parity symmetry. To this end, we consider a simple case in which the symmetry is broken by taking the two coupling capacitors to have different values:
\be
C_c^0 = C_c + C_t \delta c/2 \, , \qquad C_c^N = C_c - C_t \delta c/2
\ee
with $C_t$ defined in \eref{Ctdef}.
Therefore in this section $C_c$ represent the average of the two coupling capacitors, and we have introduced the dimensionless asymmetry parameter
\be
\delta c = \frac{C_c^0 - C_c^N}{C_t}
\ee
with $|\delta c| \le 1$. Inclusion of the asymmetric capacitive coupling in the circuit Lagrangian  amounts to the substitutions \cite{fasy}
\bea
G_{00} & \to & G_{00} -\frac{1}{4E_t}\left(\delta c\right)^2  \, ,\label{G00s} \\
G_{0\mu} & \to & G_{0\mu} + \frac{\delta c}{2E_g^a}\frac{c_\mu o_\mu}{\sqrt{2N} s^2_\mu} \label{G0ms} \\
\eea
in \esref{G00}-\rref{G0m}, and
\be\label{LVs}
{\cal L}_V \to {\cal L}_V + \frac{\left(\delta c\right)^2}{8 E_t } \dot{\phi} eV - \frac{\delta c}{4E_g^a} \sum_\mu \frac{c_\mu o_\mu}{ \sqrt{2N} s^2_\mu} \dot{\xi}_\mu eV
\ee
in \eref{LV1}. Note that matrix $G_{\mu\nu}$ is not affected and, as discussed in Sec.~\ref{sec:odd}, a rotation is needed to diagonalize it in the odd sector. This rotation in principle modifies the new term introduced in \eref{G0ms}; we neglect this modification as it does not introduce any new qualitative feature -- this is also a quantitatively good approximation if the parameter $\lambda$, \eref{lam_def}, is sufficiently small.

Within the same approximations used previously [in particular, we assume again \eref{Hcon} to hold], the total Hamiltonian $H$ takes the same form as in \eref{He}:
\bea
H & = & \bar{H}_\phi + \sum_{\mu=1}^{N-1} H_\mu + H_\mathrm{int} + H_V \, , \label{Htot} \\
\bar{H}_\phi & = & 4\bar{E}_C^\phi p_\phi^2 - E_J^b \cos \left(\phi + \varphi_e\right) + \frac12 E_L \phi^2 \, ,\label{Hphib} \\
H_\mu & = & 4E_{C,\mu} p_\mu^2 +\frac12 E_J^a \xi_\mu^2 \, , \label{Hmu}\\
H_\mathrm{int} & = & \sum_{\mu=1}^{N-1} g_\mu p_\mu p_\phi \, , \\
H_V & = & - \sum_{\mu=1}^{N-1}  g_\mu p_\mu \,eV \left[ \frac{1}{4E_C^c} - \frac{(\delta c)^2}{8E_t} - \frac{o_\mu}{2\bar{E}_C^\phi} \right]  \label{HVa}\\
&-& \bar{E}_C^\phi p_\phi \, eV  \left[\frac{2}{E_C^c}-\frac{(\delta c)^2}{E_t} + \frac{(\delta c)^2}{2N}\sum_{\mu} o_\mu \frac{E_{C,\mu}}{\left(E_g^a\right)^2}
\frac{c_\mu^2}{s_\mu^4} \right]. \nonumber
\eea
Despite the formal similarity, there are important differences between \esref{He} and \rref{Htot}: first, all collective modes appear in $H$, not just the even ones; in fact, we remind here that $E_{C,\mu}$ is given by either \eref{ecodd} or \eref{eceven} depending on the mode parity. Second, due to the asymmetry the qubit charging energy $\bar{E}_C^\phi$ is renormalized from the definition in \eref{tEcp}:
\be
\frac{1}{\bar{E}_C^\phi} = \frac{1}{\tilde{E}_C^\phi} - \frac{(\delta c)^2}{4E_t} \, .
\ee
Third, the coupling constants $g_\mu$ are different for even and odd modes:
\be
\quad g_\mu = \frac{4}{\sqrt{2N}} \frac{\bar{E}_C^\phi E_{C,\mu}}{E_g^a} \frac{c_{\mu}}{s_{\mu}^2} \left(o_{\mu+1} - o_\mu \delta c\right) \, .
\ee

The structure of Hamiltonian $H$ in \eref{Htot} shows that the main consequence of breaking the parity symmetry is the introduction of coupling between qubit and odd modes with coupling strength linear in the asymmetry parameter $\delta c$. Therefore for strong asymmetry, $|\delta c| \sim 1$, the odd modes influence the qubit in the same way as the even ones. Even for moderate asymmetry, $|\delta c| \sim 0.1$, the effect of the lower-energy odd modes may be non-negligible (at least near zero flux, where the qubit frequency is closer to those of the collective modes): while $\delta c$ suppresses the coupling of the odd modes to the qubit, the odd modes with index $2\rho-1$ are closer in frequency to the qubit than the even modes with index $2\rho$, and the smaller frequency difference generally increases the dispersive shift, see \eref{chir}; also, the term proportional to $1/\bar{E}_C^\phi$ in \eref{HVa} roughly compensate for the $\delta c$ suppression of coupling between odd modes and cavity, thus giving similar lifetimes for odd ($2\rho-1$) and even ($2\rho$) modes. On the other hand, small asymmetry at the percent level, as usually present in nominally symmetric devices, implies that the odd modes can be safely neglected.

\subsection{Comparison with experiment}
\label{sec:expcomp}

Asymmetrically coupled systems similar to that described above have been recently probed experimentally, which enable us to test in part our theory. For example, in \ocite{supind} an array of 80 junctions was placed in parallel to a (resonator) capacitor and the nine lowest resonant frequencies were measured. The system is described by the Hamiltonian $H$ in \eref{Htot} if we set $E_J^b = 0$. Then $H$ describes a harmonic oscillator linearly coupled to $N-1$ oscillators. The resonant frequencies of the corresponding $N$ independent oscillators can be easily calculated numerically; to compare with experiments, we note that the lowest mode in the experiment correspond to what we call the qubit mode $\phi$, and therefore the higher even indices correspond to our odd modes and viceversa, odd mode indices in the experiments correspond to our even modes.

\begin{figure}[bt]
 \includegraphics[width=0.48\textwidth]{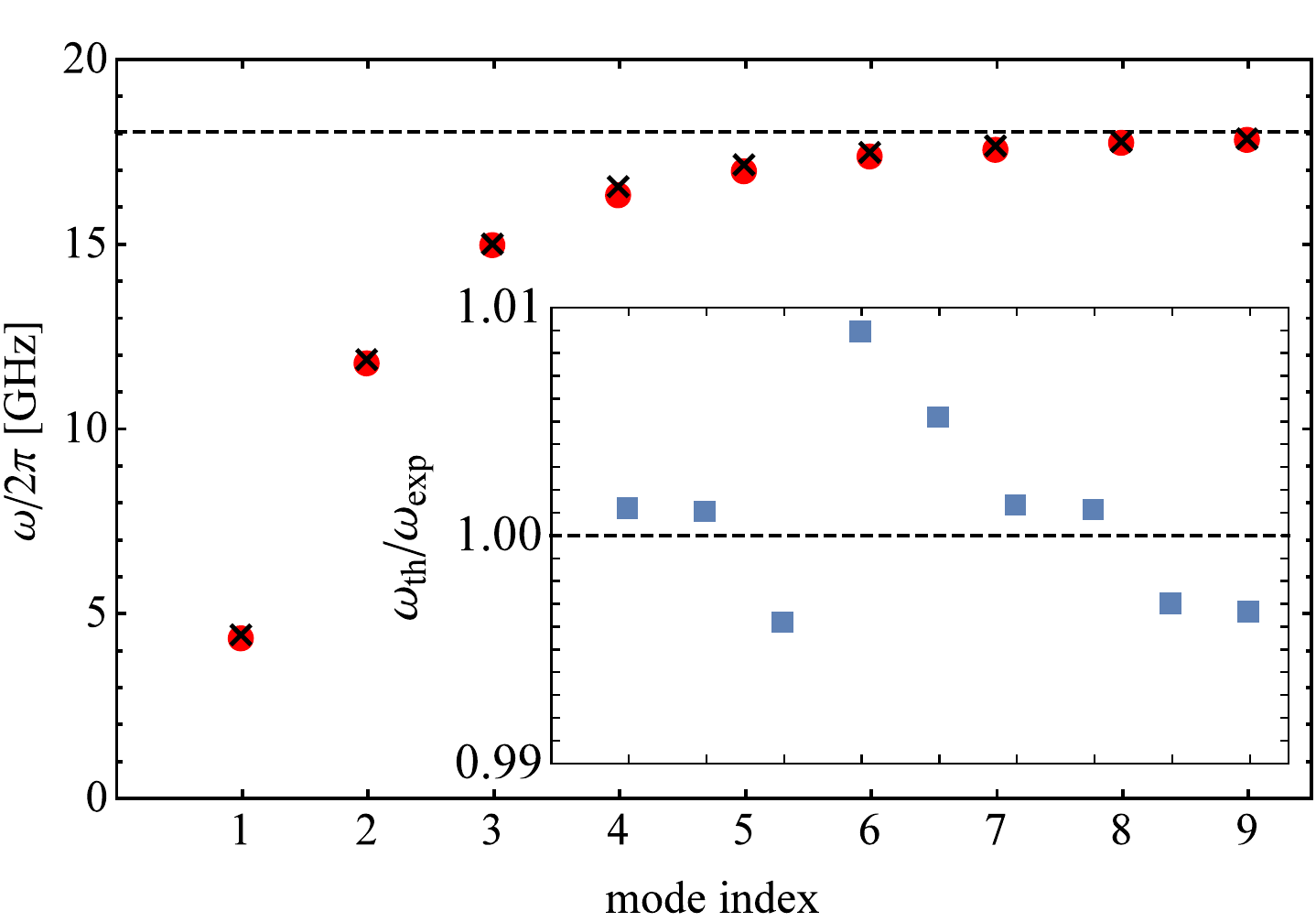}
 \caption{(Color online) Comparison between collective modes frequencies measured in \ocite{supind} (red circles) with those calculated using the present theory (black crosses). The inset shows the ratio between theoretical and experimental frequencies.}
 \label{expfit}
\end{figure}

In calculating the resonant frequencies we use as input parameters the array junction capacitance $C_J^a=40.1$~fF ($E_C^a \simeq 483$~MHz) and its Josephson inductance $L_J = 1.94$~nH ($E_J^a \simeq 84.3$~GHz), the capacitance to ground $C_g^a = 114$~aF ($E_g^a \simeq 170$~GHz), the two coupling capacitors $C_c^0 = 5.5$~fF and $C_c^N = 3.16$~fF ($E_C^c \simeq 2.24$~GHz, $\delta c \simeq 0.27$), and the resonator capacitor $C_J^b = 3.19$~fF ($E_C^b \simeq 6.07$~GHz). The frequencies calculated with these parameters differ by less than 1\% from the measured frequencies, see Fig.~\ref{expfit}. We note that while the array junction parameters $C_J^a$ and $L_J$ agree with those reported in \ocite{supind}, the ground capacitance $C_g^a$ we estimate here is about three times bigger; since the calculations in the previous sections are based on the original estimate of \ocite{supind}, they could underestimate, \textit{e.g.}, the dispersive shifts by almost one order of magnitude.

A very recent experiment \cite{wg} reports the measurement of 14 resonant frequencies in an array of 200 junctions without shunting capacitor ($C_J^b = 0$). We can again compare our calculated frequencies with the measured ones: setting $C_g^a = 98$~aF and optimizing the other parameters, we find again differences of less than 1\% except for the third mode, whose measured frequency is about 7\% higher than the calculated one; this larger difference is likely due to the presence near the frequency of that mode of a spurious resonance \cite{wg}.

\section{Coupling into the superinductance}
\label{sec:sic}

So far, both for the $PT$-symmetric and the broken-symmetry cases, we have taken the coupling capacitors to be connected to the two islands separated by the phase-slip junction. The coupling capacitors can be attached to any island in the circuit, and in fact such a setup has been used in more recent experiments \cite{nature,jumps}. In general, arbitrary placement will immediately break parity symmetry even if the capacitances are the same for both capacitors. Here we consider briefly the simplest case of equal capacitors placed symmetrically with respect to the phase-slip junction, so that parity symmetry is preserved and no qubit-odd mode interaction is allowed. Concretely, we take the first capacitor to be connected to island $\delta \ge 0$, where islands $0$ and $N$ are the two island surrounding the phase-slip junction; then the second capacitor is connected to island $N-\delta$, and the maximum possible $\delta$ is $\delta_M = \lfloor (N-1)/2 \rfloor$. In this configuration, the coupling Lagrangian ${\cal L}_V$ [cf. \eref{LV1}] becomes
\be\label{LV2}
{\cal L}_V = - \frac{1}{4E^c_C}\left(1-\frac{2\delta}{N}\right)\dot\phi\, e V + \frac{1}{4E^c_C} \sqrt{\frac2N}
\sum_{\rho=1}^{N_e}\frac{s_{4\delta\rho}}{ s_{2\rho}}\dot\eta_\rho \, e V \, .
\ee
This formula correctly reduces to \eref{LV1} for $\delta =0$, while for $\delta>0$ a new coupling between cavity and even modes is present.

Changing the position of the coupling capacitors also affects the kinetic energy part $T_G$ of the Lagrangian [cf. \eref{TG}], and a general treatment of this modified term is quite cumbersome. Here we consider the simple limit in which we neglect the ground capacitances, $C_g^a, C_g^b \to 0$.
In this case, as we show in Appendix~\ref{app:hcs}, the Hamiltonian is
\bea
H & = & \hat{H}_\phi + H_2 + H_\mathrm{int} + H_V + \sum_{\mu\neq 2} H_\mu \, , \label{hcs} \\
\hat{H}_\phi & = & 4\frac{\hat{E}_C^\phi}{1-\mathrm{g}^2} p_\phi^2 - E_J^b \cos \left(\phi + \varphi_e\right) + \frac12 E_L \phi^2  \, ,\\
H_2 & = & 4\frac{E_{C2}}{1-\mathrm{g}^2} p_2^2 +\frac12 E_J^a \xi_2^2 \, ,\\
H_\mathrm{int} & = & \frac{\mathrm{g}\sqrt{\hat{E}_C^\phi E_{C2}}}{1-\mathrm{g}^2} p_2 p_\phi \equiv \mathrm{g}_2  p_2 p_\phi \, , \label{hint2} \\
H_\mu & = & 4E_{C}^a p_\mu^2 +\frac12 E_J^a \xi_\mu^2 \, ,
\eea
\be\label{hvs}\begin{split}
H_V  & =  -\frac{2 p_\phi eV}{1-\mathrm{g}^2} \frac{\hat{E}_C^\phi}{E_C^c} \left(1-\frac{2\delta}{N}\right)\left[1-\delta\left(1-\frac{2\delta}{N}\right)\frac{E_{C2}}{E_C^c}\right] \\
+ & \frac{4 p_2 eV}{1-\mathrm{g}^2} \frac{E_{C2}}{E_C^c} \sqrt{\delta\left(1-\frac{2\delta}{N}\right)}
\left[1-\left(1-\frac{2\delta}{N}\right)^2\frac{\hat{E}_C^\phi}{2E_C^c}\right] \\
& \equiv -\mathrm{g}_{\phi,c} p_\phi eV + \mathrm{g}_{2,c} p_2 eV \, ,
\end{split}\ee
where the parameters $\mathrm{g}_2$, $\mathrm{g}_{\phi,c}$, and $\mathrm{g}_{2,c}$ defined above as well as
\bea
\frac{1}{\hat{E}_C^\phi} & = & \frac{1}{E_C^\phi} + \frac{1}{2E_C^c} \left(1-\frac{2\delta}{N}\right)^2 , \\
\frac{1}{E_{C2}} & = & \frac{1}{E_C^a} + \frac{1}{E_C^c} \delta \left(1-\frac{2\delta}{N}\right) \, , \\
\mathrm{g}^2 & = & \frac{\hat{E}_C^\phi E_{C2}}{2\left(E_C^c\right)^2} \delta \left(1-\frac{2\delta}{N}\right)^3 \label{g2def}
\eea
depend on the position $\delta$ of the coupling capacitors.

\begin{figure}[bt]
 \includegraphics[width=0.483\textwidth]{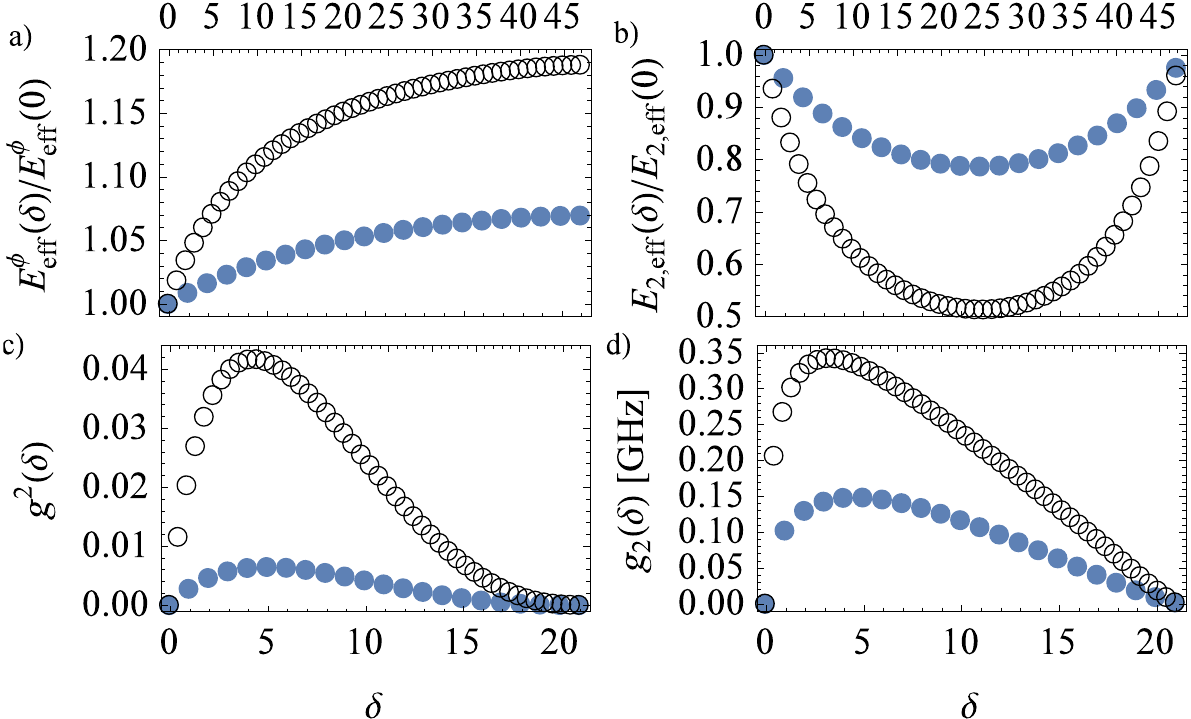}
 \caption{(Color online) In all four panels, filled circles are used to denote parameter set 1 and empty ones for set 2. The horizontal bottom (top) scales are used for set 1 (2). a) Normalized effective qubit charging energy $E_\mathrm{eff}^\phi(\delta)/E_\mathrm{eff}^\phi(0)$ vs. coupling capacitor position $\delta$. b) Normalized collective mode charging energy $E_{2,\mathrm{eff}}(\delta)/E_{2,\mathrm{eff}}(0)$ vs. $\delta$. c) dimensionless parameter $\mathrm{g}$ [\eref{g2def}] vs. $\delta$. d) qubit-collective mode coupling $\mathrm{g}_2$ [\eref{hint2}] vs. $\delta$.}
 \label{hspf}
\end{figure}

The main qualitative feature of the Hamiltonian in \eref{hcs} is that the qubit and the cavity both couple to only one collective mode whose charging energy is renormalized below $E_C^a$, while all the other $N-2$ modes remain degenerate and uncoupled. (Of course in the presence of ground capacitances the degeneracy is lifted and all the even modes couple to both qubit and cavity.) While for typical experimental parameters the effective qubit charging energy $E^\phi_{\mathrm{eff}} = \hat{E}_C^\phi/(1-\mathrm{g}^2)$ moderately increases as  $\delta$ increases towards $N/2$, the collective mode effective charging energy $E_{2,\mathrm{eff}} = E_{C2}/(1-\mathrm{g}^2)$ can be more strongly suppressed when $\delta \sim N/4$, see Figs.~\ref{hspf}a and \ref{hspf}b. The dimensionless parameter $g$ varies non-monotonically as function of $\delta$ and is generally small, see Fig.~\ref{hspf}c. The qubit-collective mode coupling strength $\mathrm{g}_2$ also depends significantly on $\delta$, see Fig.~\ref{hspf}d; note that the largest values of $\mathrm{g}_2$ at $\delta \sim N/8$ are a significant fraction of (or comparable to) the coupling strengths between qubit and lowest collective modes calculated for $\delta =0$ but in the presence of ground capacitances, see Fig.~\ref{grfig}. These observations imply that by appropriately placing additional capacitors in the array, both the collective modes spectrum and the coupling strength with the qubit can be controlled to some degree. In particular, to minimize the effects of the collective modes on the qubit the coupling capacitors should either be placed next to the phase-slip junction ($\delta = 0$), or opposite to it ($\delta = \delta_M$), while intermediate positions (especially in the range $\delta \sim N/8 - N/4$) maximize those effects.

\begin{figure}[tb]
 \includegraphics[width=0.48\textwidth]{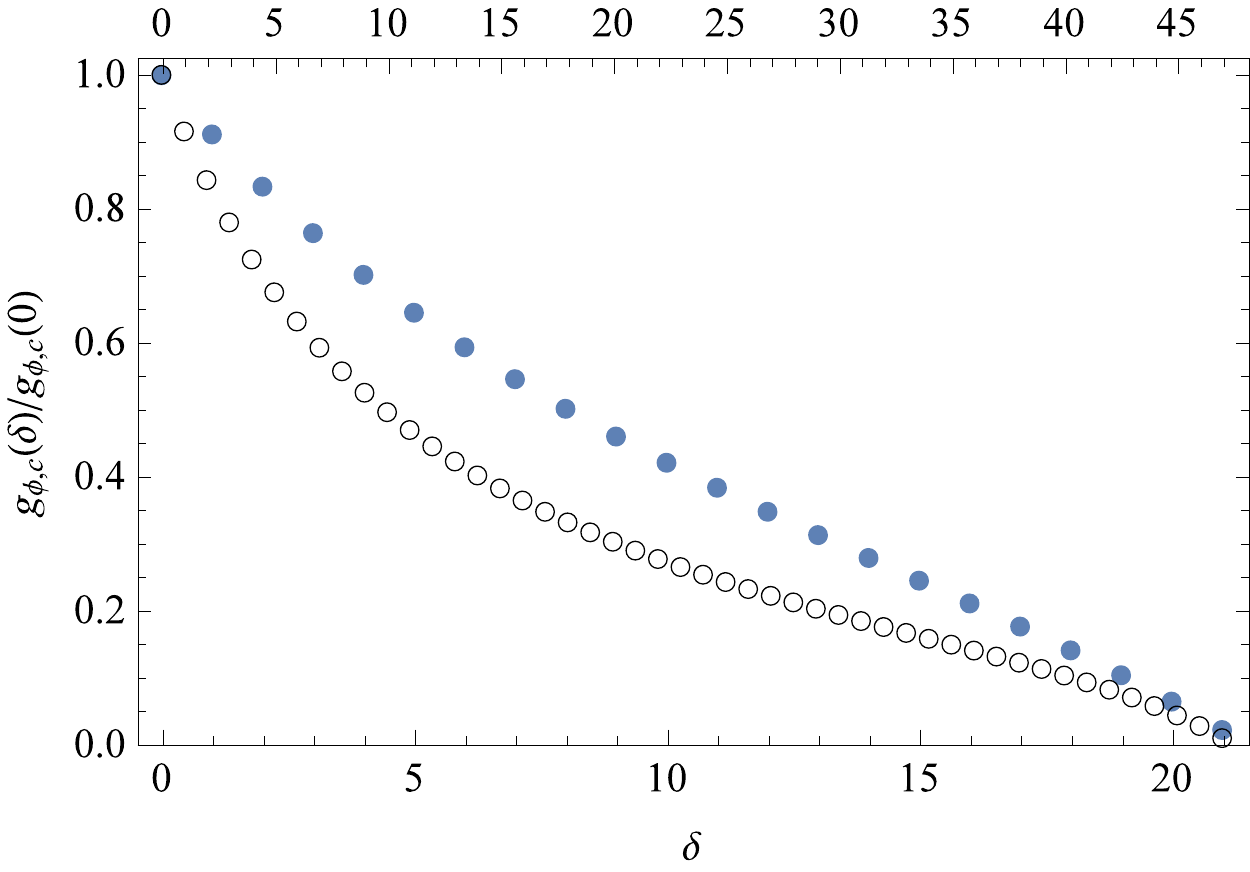}
 \caption{(Color online) Normalized qubit-cavity coupling $\mathrm{g}_{\phi,c}$ [\eref{hvs}] vs. coupling capacitors position $\delta$. Horizontal bottom (top) scale is used for set 1, filled circles (2, empty circles).}
 \label{gfc}
\end{figure}

\begin{figure}[bt]
 \includegraphics[width=0.48\textwidth]{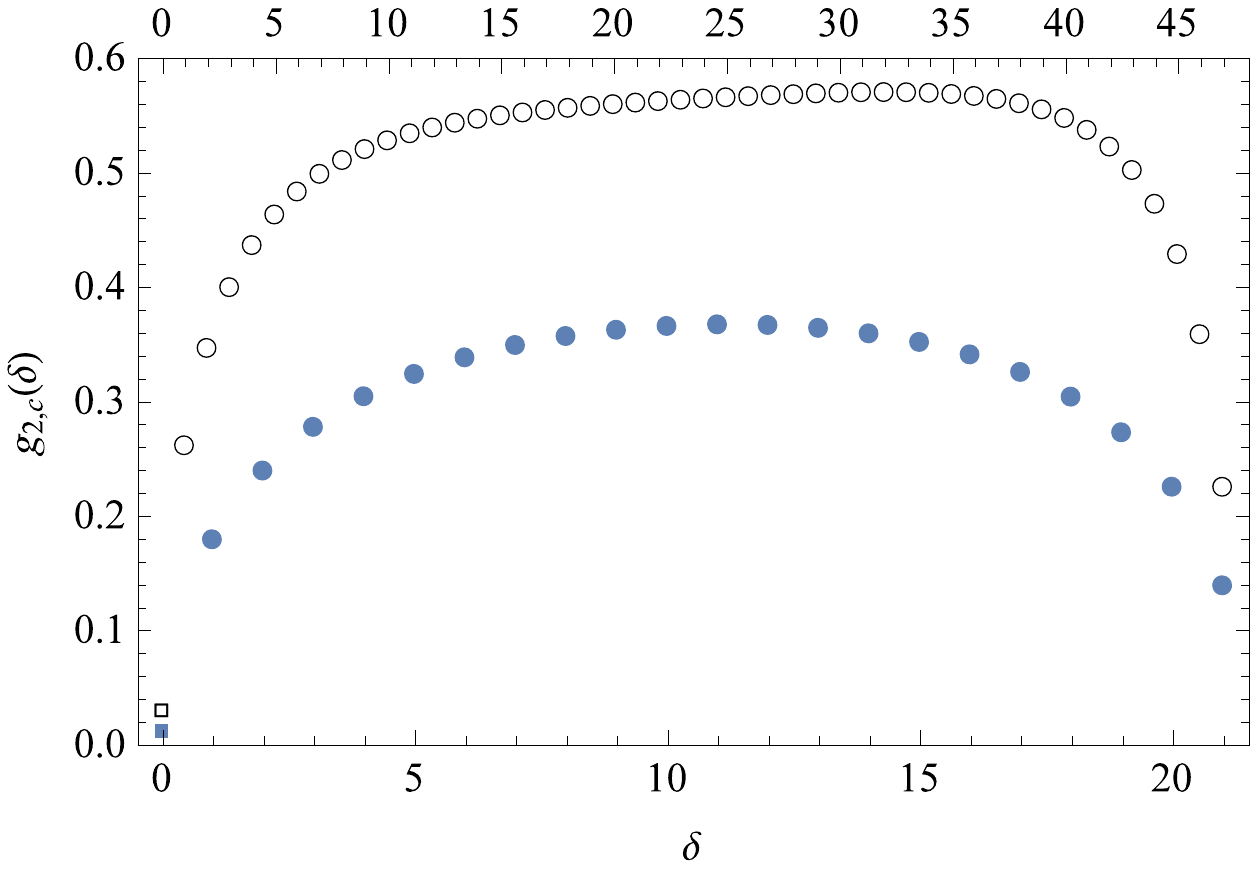}
 \caption{(Color online) Collective mode-cavity coupling $\mathrm{g}_{2,c}$ [\eref{hvs}] vs. coupling capacitors position $\delta$. Horizontal bottom (top) scale is used for set 1, filled circles (2, empty circles). The squares at $\delta=0$ are given by $g_1/4E_C^c$ [cf. \eref{HV}], with $g_1$ from \eref{Hint}, and are plotted here for comparison.}
 \label{g2c}
\end{figure}

In \eref{hvs} we give expressions for the qubit-cavity and collective mode-cavity dimensionless couplings $\mathrm{g}_{\phi,c}$ and $\mathrm{g}_{2,c}$. For $\delta=0$, $\mathrm{g}_{\phi,c}$ reduces to the value in \eref{HV}; as $\delta$ increases, it gradually decreases down to a value approximately $1/N$ times the initial one when $\delta=\delta_M$. In contrast, $\mathrm{g}_{2,c}$ takes its smallest values for $\delta=1$ and $\delta = \delta_M$, where it is approximately given by $\mathrm{g}_{2,c}^{\mathrm{min}} \approx 4E_C^a/\sqrt{2}E_C^c$; the largest values at $\delta \sim N/4$ are bigger by a factor of less than 3. It turns out [cf. Fig.~\ref{g2c}] that for typical experimental parameters even the minimum value $\mathrm{g}_{2,c}^{\mathrm{min}}$ is larger than the strongest collective mode-cavity coupling $g_1/4E_C^c$ in \eref{HV} \cite{ffail}.

The contrasting dependence on $\delta$ of the two couplings indicates that moving the coupling capacitors away from the phase-slip junction can adversely affect the qubit coherence: indeed as $\delta$ increases, the coupling capacitors must also increase to attain the desired qubit-cavity coupling strength; both moving the capacitors and increasing their capacitance, however, raise the collective mode-cavity coupling, which in turns increases the collective mode (Purcell) decay rate $\kappa$ (cf. Sec.~\ref{sec:purc}) and hence the qubit dephasing rate, see Sec.~\ref{sec:coher}. We therefore conclude that placing the coupling capacitors beside the phase-slip junction, $\delta=0$, is the optimal choice.

If the coupling capacitors are placed opposite to the phase-slip junction ($\delta\sim N/2$), the coupling capacitance should be increased by a factor of order $N$ to compensate for the decrease in qubit-cavity coupling strength. Together with the stronger mode-cavity coupling, $\mathrm{g}_{2,c}^{\mathrm{min}}$ as compared to $g_1/4E_C^c$, this increase would raise the collective modes decay rate by at least $\sim 10 N^2$ (i.e., $\sim10^4$). Then the decay rate of the lowest even mode would be faster than that of the qubit, the dephasing rate $\Gamma_\phi^\kappa$ of \eref{gpk} would also increase by about 4 orders of magnitudes, and the corresponding dephasing time would be about $0.1\,$ms. This time is one order of magnitude longer than the coherence time measured in \ocite{nature}, indicating that the collective modes are not limiting coherence in current experiments. On the other hand, our estimate is much shorter than the measured $\sim 10\,$ms relaxation time at half flux quantum, so the effect of the collective mode could in principle be observable, if other dephasing mechanisms can be identified and suppressed.

\section{Summary}
\label{sec:summ}

In this paper we study the collective modes in the array of Josephson junction forming the superinductance of the fluxonium qubit. We derive an approximate Hamiltonian, \eref{He}, that includes the interactions between the qubit mode and the collective modes in the presence of ground capacitances. The approximations place some restriction on the number of array junction to which the model applies, see \eref{Hcon}, but this condition is in practice much weaker than that given by the array ``screening length'' [\eref{smallN}] and it is satisfied in current experiments. A generalization of this Hamiltonian enable us to favorably compare the calculated spectrum of the collective modes to two recent experiments, see Sec.~\ref{sec:sb} and Fig.~\ref{expfit}.

In Sec.~\ref{sec:anhar} we consider the leading-order non-linearity of the array junctions, which introduces additional qubit-collective mode interactions. Among these interactions, the term which leads to the strongest dispersive shifts effectively induces fluctuations in the qubit inductive energy when the collective modes are excited, see Sec.~\ref{sec:u2}. As we discuss in Sec.~\ref{sec:coher}, the total dispersive shifts (i.e., including also the effect of ground capacitances) are much bigger than the collective mode decay rates, so the latter determine the qubit dephasing rate. We find that the collective modes do not significantly contribute to dephasing, so long as they are more weakly coupled to the cavity than the qubit is; the weak coupling is generically achieved if the qubit-cavity coupling capacitors are placed next to the phase-slip junction. However, we estimate in Sec.~\ref{sec:sic} that the collective-mode induced dephasing could become observable if the coupling capacitors are placed opposite to the phase-slip junction.

\acknowledgments

We gratefully acknowledge interesting discussions with D. DiVincenzo, F. Kons\c{c}helle, S. Mehl, I. Pop, G. Rastelli, F. Solgun, and U. Vool. This work was supported in part by the Alexander von Humboldt and Knut och Alice Wallenbergs foundations (GV) and by the EU under REA Grant Agreement No. CIG-618258 (GC).

\appendix

\section{Lagrangian}
\label{app:lder}

This appendix provides a derivation of the Lagrangian ${\cal L}$ for the collective and qubit modes [\eref{Ltot}] starting from a more familiar ``textbook'' formulation in terms
of phases and voltages. To this end, we split ${\cal L}$ as a sum of kinetic and potential energy parts as usual, ${\cal L} = T - U_S$, and write the potential energy as
\be
U_S = - E_J^a \sum_{m=1}^N \cos \theta_m -E_J^b \cos \left(\sum_{m=1}^N \theta_m + \varphi_e \right).
\ee
Here $\theta_m$ is the (gauge-invariant) phase difference across junction $m$ in the array, and the first term on the right hand side is the Josephson energy of the array junctions. The last term in the above equation is the phase-slip junction energy, and in writing this term we have taken into account the fluxoid quantization condition
\be\label{fqc}
\sum_{m=0}^N \theta_m + \varphi_e = 2\pi n
\ee
with $n$ integer.

The kinetic energy part $T$ is more easily expressed in terms of the voltages $\dot\varphi_m /2e$ of each island:
\bea
T &=& T_S + T_G + T_V \, ,\\
T_S &=& \sum_{m=1}^{N}\frac{\left(\dot\varphi_m - \dot\varphi_{m-1} \right)^2}{16E_C^a}  + \frac{\left(\dot\varphi_N - \dot\varphi_{0} \right)^2}{16E_C^b} \, , \\
T_G &=& \sum_{m=0}^{N}\frac{\dot{\varphi}_m^2}{16E_g^m} \, , \\
T_V &=& \sum_{m=0}^{N}\frac{\left(\dot{\varphi}_m - 2e V_m \right)^2}{16E_c^m} \, ,
\eea
Here $T_S$ is the charging energy due to the junctions capacitances, $T_G$ due to capacitances between each island and ground, and $T_V$ due to capacitive coupling to external
voltage sources $V_m$. In the above equations $E_g^m = e^2/2C_g^m$ and $E_c^m=e^2/2C_c^m$ are charging energies of ground and coupling capacitors for the $m$th island and they can in general be different for each island. We stress that all equations in this Appendix are valid for this generic case, not just for the specific circuit depicted in Fig.~\ref{flu1_b}.

To rewrite $T$ in terms of the phase differences $\theta_m$, we use the relationship
\be
\varphi_m = \varphi_0 + \sum_{l=1}^m \theta_l \, ,
\ee
valid for $m=1,\ldots, N$, where we have taken $\varphi_0$ as a reference phase. Then it is straightforward to write $T_S$ in terms of $\theta_m$ variables
\be
T_S = \sum_{m=1}^{N}\frac{\dot{\theta}_m^2}{16E_C^a}  + \frac{\left(\sum_{m=1}^N \dot\theta_m \right)^2}{16E_C^b} \, .
\ee
The other two terms in $T$ take the form
\bea
T_G &=& \frac{\dot{\varphi}_0^{2}}{16E_{g}^{0}}
+\sum_{m=1}^{N} \frac{1}{16E_{g}^{m}}\left(\dot{\varphi}_0+\sum_{l=1}^{m}\dot{\theta}_{l}\right)^{2}  \, ,\\
T_V &=& \frac{\left(\dot{\varphi}_0-2eV_0\right)^{2}}{16E_{c}^{0}} \\ & &
+\sum_{m=1}^{N} \frac{1}{16E_{c}^{m}}\left(\dot{\varphi}_0+\sum_{l=1}^{m}\dot{\theta}_{l}-2eV_m\right)^{2}. \nonumber
\eea

We next note that ${\cal L}$ is independent of $\varphi_0$, so that $\partial {\cal L}/\partial \dot\varphi_0$ is a conserved quantity, the total charge of the circuit \cite{prx}. Using this conservation law we can express $\dot\varphi_0$ in terms of the variables $\theta_m$ and thus eliminate it from the Lagrangian. In this way, standard algebraic manipulations lead to
\bea
T_G &=& \frac{1}{16} \sum_{m,n=1}^{N} {\cal G}_{mn} \dot\theta_m \dot\theta_n \, , \\
{\cal G}_{nm} &=& E_t \!\!\!\!\sum_{i=0}^{\min\{m,n\}-1}\!\!\!\sum_{j=\max\{m,n\}}^N\!\left(\frac{1}{E_g^i}+\frac{1}{E_c^i}\right)\!\left(\frac{1}{E_g^j}+\frac{1}{E_c^j}\right) , \nonumber
\\
T_V &=& \frac{2e}{8} \sum_{m=1}^N \dot\theta_m \sum_{i=m}^N \left[ \left(\frac{1}{E_g^i}+\frac{1}{E_c^i}\right) \bar{V} - \frac{V_i}{E_c^i}\right] , \\
\bar{V} & = & E_t \sum_{i=0}^N \frac{V_i}{E_c^i} \, , \nonumber
\eea
where
\be
\frac{1}{E_t} = \sum_{i=0}^N \left(\frac{1}{E_g^i}+\frac{1}{E_c^i}\right) .
\ee
These equations correctly reduce to those of \ocite{prx} in the absence of coupling capacitors.

As a final step, we introduce a new set of variables via the relations
\bea
\phi &=& \sum_{m=1}^N \theta_m \, , \\
\xi_\mu &=& \sum_{m=1}^N W_{\mu m} \theta_m
\eea
with index $\mu=1,\, \ldots, N-1$ and inverse $\theta_m = \phi/N + \sum_\mu W_{\mu m} \xi_\mu$. The matrix $W_{\mu m}$ must satisfy the conditions $\sum_m W_{\mu m} = 0$ and $\sum_m W_{\mu m} W_{\nu m} = \delta_{\mu\nu}$. In terms of these new variables we find $T_S$ as in \eref{TS} and $U_S$ as in \eref{US}. Formulas for $T_G$ and $T_V$ and arbitrary $W_{\mu m}$ are not instructive, so we do not report them here. For the specific circuit configuration and choice of $W_{\mu m}$ described in Sec.~\ref{sec:model}, the corresponding formulas are given there and follow directly from the equations above. Modifications of those formulas for a different circuit configuration breaking parity symmetry are discussed in Sec.~\ref{sec:sb}. A third circuit with coupling capacitors connected into the array is briefly considered in Sec.~\ref{sec:sic}. Here we mention a useful identity valid for the choice of $W_{\mu m}$ in \eref{wmn}:
\be
\sum_m m W_{\mu m} = - \frac{1}{\sqrt{2N}} \frac{c_\mu}{s^2_\mu} o_\mu
\ee
[for the notation used, see the definitions in \esref{sm}-\rref{om}].

\section{Choice of parameters}
\label{app:par}

An often measured property of a flux-tunable qubit such as the fluxonium is its spectrum as a function of flux, $\EZO(f)$. The spectrum can be obtained by numerical diagonalization of the qubit Hamiltonian $H_\phi$, \eref{Hphi}, where the inductive energy $E_L$ should be replaced by $\tilde{E}_L$ [\eref{tEl}]. For the experiments reported in Refs.~\onlinecite{flux1} and \onlinecite{nature}, with $N=43$ and $N=95$ array junctions respectively, this procedure leads to the parameters reported in Table~\ref{taba}. We also give there our rough estimate of the ratio $C_g^b/C_g^a$ which is based on the geometry of the devices in the two experiments.

The phase-slip junction Josephson energy $E_J^b$ in Table~\ref{tabpar} is taken directly from the experimental estimates in Table~\ref{taba}. For the coupling capacitor energy $E_C^c$, we use for set 1 the value of coupling capacitance given in \ocite{flux1}, while for set 2 we take as an example the value reported in \ocite{supind}. The latter experiment was performed with junctions fabricated with the same procedure used to fabricate the fluxonium junctions of \ocite{nature}; for this reason, we also use the value of ground capacitance in \ocite{supind} to estimate $E_g^a$ for set 2. The value of $E_g^a$ for set 1 is smaller than that of set 2 by a factor of 0.4 because, according to the supplementary to \ocite{supind}, the different fabrication processes lead to such a difference in capacitances to ground. Values for $E_g^b$ then follows from the geometrically estimated ratio $C_g^b/C_g^a$. To further constraint the parameters, since the only difference between phase slip and array junctions is in their area, we assume that their plasma frequency is the same, so that $E_J^a E_C^a = E_J^b E_C^b$. We then choose the parameters $E_J^a$, $E_C^a$, and $E_C^b$ as to obtain, using \esref{tEcp} and \rref{tEl}, the values of $\tilde{E}^\phi_C$ and $\tilde{E}_L$ reported in Table~\ref{taba}.

\begin{table}[b]
\begin{tabular}{|c||c|c|c|c||c|}
\hline
& $N$ & $E_J^b$ & $\tilde{E}_C^\phi$ & $\tilde{E}_L$ & $C_g^b/C_g^a$ \\ \hline
set 1 & 43 & 8.93 & 2.39 & 0.52 & 32 \\
set 2 & 95 & 10.2 & 3.60 & 0.46 & 0.6 \\
\hline
\end{tabular}
\caption{Additional fluxonium parameters estimated as explained in Appendix~\ref{app:par}. Energies are given in GHz.}
\label{taba}
\end{table}

\section{Rotation in the odd sector}
\label{app:rot}

The diagonalization of the kinetic energy matrix $T_o$ [\eref{To0}] of the odd sector can be seen as a rotation among the odd-sector coordinates $\zeta_\rho$. Such a rotation
can be constructed perturbatively order by order in the parameter $\lambda$, with a procedure analogous to that used to perform a Schrieffer-Wolff transformation to an effective Hamiltonian: we want to obtain an antisymmetric matrix $S = \sum_{n=1} S_n \lambda^n$ such that the the product $e^{-S} T_o e^S$ is diagonal. Expanding this product up to second order we have
\be\begin{split}
& e^{-S} T_o e^S  = T_d + \lambda \left(T_\lambda - \left[S_1, T_d \right] \right)  \\ & + \lambda^2 \left(\frac12 \left[S_1, \left[S_1,T_d \right]\right] - \left[S_2, T_d\right] - \left[S_1, T_\lambda\right] \right) + \ldots
\end{split}\ee
To eliminate the off-diagonal terms at order $\lambda$, we take $S_1$ to have elements $S_{1,\rho\rho}=0$ and for $\rho\neq\sigma$
\be
S_{1,\rho\sigma} = \frac{T_{\lambda,\rho\sigma}}{T_{d,\sigma\sigma} - T_{d,\rho\rho}}\, ,
\ee
where for example the elements $T_{\lambda,\rho\sigma}$ of matrix $T_\lambda$ are defined via $T_\lambda = \sum_{\rho,\sigma} T_{\lambda,\rho\sigma}\dot\zeta_\rho \dot\zeta_\sigma$. We can similarly choose $S_2$ to eliminate the off-diagonal terms at order $\lambda^2$. The remaining diagonal elements are then, up to order $\lambda^2$,
\be\label{Todia}
T_{d,\rho\rho} + \lambda T_{\lambda,\rho\rho} - \frac{\lambda^2}{2} \left[S_1, T_\lambda \right]_{\rho\rho} \, .
\ee
The last term is more explicitly written as
\be
\left[S_1, T_\lambda \right]_{\rho\rho} = 2\sum_{\sigma\neq\rho}  \frac{T_{\lambda,\rho\sigma}T_{\lambda,\sigma\rho}}{T_{d,\sigma\sigma} - T_{d,\rho\rho}} \, .
\ee
By substituting the matrix elements into \eref{Todia}, we arrive at \eref{ecodd}.

Note that the rotated modes $\tilde\zeta_\rho$ are related to the original modes $\zeta_\sigma$ via $\tilde\zeta_\rho = \left(e^S\right)_{\rho\sigma} \zeta_\sigma$.
For use in Sec.~\ref{sec:effpot}, we define the matrix $\Lambda_{\mu\nu}$ that performs the rotation in the odd sector while leaving the even sector unchanged:
\be\label{LMdef}
\Lambda_{\mu\nu} = \left\{
\begin{array}{ll}
\left(e^S\right)_{2\mu-1,2\nu-1} \phantom{m} & \mu,\nu \, \mathrm{odd} \\
\delta_{\mu\nu} & \mu,\nu \, \mathrm{even} \\
0 & \mathrm{otherwise}
\end{array}\right.
\ee

\section{Even sector Hamiltonian}
\label{app:He}

The Lagrangian ${\cal L}_e$ in \eref{Leven} has a quadratic kinetic energy part, so that transforming to the Hamiltonian amounts to inverting the $(N_e+1)\times (N_e+1)$ matrix $K= K_d + K_i$ whose diagonal part $K_d$ has entries
\bea
K_{d,00} &=& \frac{1}{8\tilde{E}_C^\phi} \, ,\\
K_{d,\rho\rho} & = & \frac{1}{8E^e_{C,\rho}}\, , \quad \rho=1,\ldots,N_e \, ,
\eea
and the interaction matrix $K_i$  has elements
\be
K_{i,0\rho} = K_{i,\rho 0} = -\frac{1}{32\sqrt{2N}} \frac{1}{E_g^a} \frac{c_{2\rho}}{s^2_{2\rho}}
\ee
and all other elements are zero. Indeed, after performing a standard Legendre transformation the kinetic part $H_K$ of the Hamiltonian has the form
\be\label{HK}
H_K = \frac12 n K^{-1} n - n K^{-1} v \, ,
\ee
where
\bea
n &=& (n_\phi, n_1, \ldots, n_{N_e}) \, , \\
v &=& (-eV/4E_C^c,0,\ldots,0)
\eea
are two $N_e+1$-dimensional vectors. We can formally write the inverse matrix as
\be\begin{split}
K^{-1} & = K_d^{-1} \frac{1}{1-K_i K_d^{-1} K_i K_d^{-1}} \\ & - K_d^{-1} K_i K_d^{-1} \frac{1}{1-K_i K_d^{-1} K_i K_d^{-1}}
\end{split}\ee
and approximate it as
\be\label{Kinvapp}
K^{-1} \approx K_d^{-1} - K_d^{-1} K_i K_d^{-1}
\ee
provided that entries of matrix $K_i K_d^{-1} K_i K_d^{-1}$ are small compared to unity. Given the structure of matrix $K_i$, this condition translates into
\be\label{Hcond}
K_{i,0\sigma}K_{i,0\rho}K_{d,\rho\rho}^{-1} K_{d,00}^{-1} = \frac{E_{C,\rho}^e \tilde{E}_C^\phi}{32N \left(E_g^{a}\right)^2}\frac{c_{2\rho}}{s^2_{2\rho}}\frac{c_{2\sigma}}{s^2_{2\sigma}} \ll 1 \, .
\ee
If the perturbative condition \eref{smallN} is satisfied, we can approximate $E_{C,\rho}^e \approx E_C^a$ -- see \eref{eceven}; using that in order of magnitude $\tilde{E}_C^\phi \sim E_C^a$, it then follows that the condition \rref{Hcond} is also satisfied. More interesting is the opposite regime in which \eref{smallN} is violated; then we can take
$E_{C,\rho}^e \approx 4 E_g^a s^2_{2\rho}$. Substituting this approximation into \eref{Hcond} and expanding the trigonometric functions for small $\rho$ and $\sigma$ (which maximizes the left hand side), we arrive at \eref{Hcon}. Using \eref{HK} and \rref{Kinvapp} we arrive at \esref{He}-\rref{HV}.

\section{Schrieffer-Wolff transformation for $U_{\phi\xi}^{(2)}$}
\label{app:SWU2}

In Sec.~\ref{sec:u2} we have argued that the the part proportional to $a_\mu^\dagger a_\mu$ of the potential term $U_{\phi\xi}^{(2)}$ gives rise to fluctuations in the inductive energy and hence to corresponding dispersive shifts; here we consider the remaining part of $U_{\phi\xi}^{(2)}$ proportional to $a_\mu^\dagger a_\mu^\dagger + a_\mu a_\mu$:
\be
U_2 = \sum_{\mu} \sum_{l,j} g_{\mu l j} |l\rangle\langle j| \left(a_\mu^\dagger a_\mu^\dagger + a_\mu a_\mu\right)\, ,
\ee
where
\be
g_{\mu l j} = \frac{E_J^a l_\mu^2}{4} \langle l | \left(1-\cos \frac{\phi}{N}\right) | j\rangle \simeq \frac{E_L l_\mu^2}{8N} \langle l | \phi^2 | j\rangle \, .
\ee

An effective qubit-collective modes Hamiltonian $H_\mathrm{eff}$ that includes the effect of this interaction term can be obtained by keeping the next-to-leading order, diagonal part of $e^{-S}H e^S$, where
\be
H = H_\phi + \sum_\mu H_\mu + U_2
\ee
with $H_\phi$ of \eref{Hphi} and $H_\mu$ of \eref{Hmu}, and
\be
S = \sum_{\mu} \sum_{l,j} g_{\mu l j} |l\rangle\langle j| \left[\frac{a^\dagger_\mu a^\dagger_\mu}{\epsilon_j - \epsilon_l -2\omega_\mu} +
\frac{a_\mu a_\mu}{\epsilon_j - \epsilon_l + 2\omega_\mu}  \right].
\ee
In this way we find
\be
H_\mathrm{eff} = H_\phi + \sum_\mu \left[ H_\mu + H_{\phi\mu}\right]
\ee
with
\be\label{Hpm}\begin{split}
H_{\phi\mu} = -\sum_{l,j} |l \rangle \langle l |  g_{\mu l j}^2   \bigg[ \frac{2(\epsilon_j - \epsilon_l)}{(\epsilon_j - \epsilon_l)^2 -(2\omega_\mu)^2 } n_\mu (n_\mu -1) \\ + \frac{4}{\epsilon_j - \epsilon_l +2\omega_\mu} \left(n_\mu + \frac12 \right) \bigg].
\end{split}\ee

When projected onto the qubit subspace, the last line above gives a small contribution to qubit frequency shift as well as to the dispersive shifts; note that due to our assumption $\omega_{10} < \omega_\mu$, the terms in the last line are always finite. To see their smallness, consider the inequalities
\be\label{sw2in}\begin{split}
\sum_j \left| \frac{g_{\mu 0 j}^2}{\epsilon_j -\epsilon_0 + 2\omega_\mu} \pm \frac{g_{\mu 1 j}^2}{\epsilon_j -\epsilon_1 + 2\omega_\mu} \right| < \sum_j \frac{g_{\mu 0 j}^2 + g_{\mu 1 j}^2}{\omega_\mu} \\
\simeq \left(\frac{E_L \ell_\mu^2}{8N}\right)^2 \frac{1}{\omega_\mu} \left[\left| \langle 1 | \phi^4 | 1\rangle\right| + \left| \langle 0 | \phi^4 | 0\rangle \right| \right]
\lesssim  \frac{\pi^4}{2} \frac{E_L \ell_\mu^2}{N^3}\, .
\end{split}\ee
In the last step we used that since state $|0\rangle$ ($|1\rangle$) is mostly localized at potential wells with minima between $0$ and $\sim \pm \pi$ ($\sim\pm2\pi$ and $\sim\pm\pi$) we have
\be\label{pmeb}
\left|\langle l | \phi^n | l \rangle \right| \lesssim \left[(l+1)\pi\right]^n \, , \qquad l=0,1 \,.
\ee
The last expression in \eref{sw2in} shows that the dispersive shift from the last line of \eref{Hpm} is much smaller than $\chi_\mu^{\delta E_L}$, see \esref{chidel} and \rref{chidelb}.

The term on the first line in \eref{Hpm} can diverge if the resonance condition $\epsilon_j - \epsilon_l - 2\omega_\mu = 0$ is met, but this divergence simply signals the breakdown of the perturbative approach when $\epsilon_j - \epsilon_l - 2\omega_\mu \approx g_{\mu l j}$. Therefore the coefficient of the largest possible contribution from the first line of \eref{Hpm} is much smaller in magnitude than $\pi^2 E_L \ell_\mu^2/2N$ (the smallness is due to the matrix element begin evaluated between one of the low-energy qubit states, $|0\rangle$ or $|1\rangle$, and a state with much higher energy). Hence we find that even near resonance this term is smaller than $\chi_\mu^{\delta E_L}$; moreover, one should keep in mind that the latter shifts the qubit frequency if any collective mode has at least one excitation, while the former changes the qubit frequency only if a specific mode (the one near resonance with a qubit transition) is excited at least twice, and the probability for the latter situation to happen is smaller by a factor $\sim\bar{n}/N$ if $\bar{n}$ is the average occupation probability -- see also the next Appendix.

\section{Occupation probabilities}
\label{app:occ}

In this Appendix we comment briefly on the assumed smallness of the occupation probabilities for the collective modes. We assume for simplicity an equilibrium probability for each mode, so that the probability $P_{n_\rho}$ of having $n_\rho$ excitations in mode $\rho$ is
\be
P_{n_\rho} = \frac{1}{1+\bar{n}_\rho} \left(\frac{\bar{n}_\rho}{1+\bar{n}_\rho}\right)^{n_\rho} ,
\ee
where $\bar{n}_\rho$ is the mode average occupation. For an order of magnitude estimate, we take the latter to be the same for all modes, $\bar{n}_\rho \equiv \bar{n}$, and small, $\bar{n} \ll 1$. Then the probability $P_0$ that none of the $N-1$ collective modes is occupied is simply
\be
P_0 = \frac{1}{\left(1+\bar{n}\right)^{N-1}} \simeq 1 - (N-1) \bar{n}
\ee
while the probabilities $P_1$ that one mode has a single excitation, $P_2$ that one mode has two excitations, and $P_{1,1}$ that two modes have one excitation are
\bea
P_1 & = & (N-1) \frac{\bar{n}}{\left(1+\bar{n}\right)^N} \simeq (N-1) \bar{n} \, ,\\
P_2 & = & (N-1) \frac{\bar{n}^2}{\left(1+\bar{n}\right)^{N+1}} \simeq (N-1) \bar{n}^2 \, , \\
P_{1,1} & = & \frac{(N-1)(N-2)}{2} \frac{\bar{n}^2}{\left(1+\bar{n}\right)^{N+1}} \\
& \simeq & \frac{(N-1)^2}{2} \bar{n}^2 \, . \nonumber
\eea
In approximating the above formulas, we assumed $\bar{n} \ll 1/N \ll 1$, and it is evident that under these assumptions we have $P_2 \ll P_{1,1} \ll P_1 \ll P_0$.

\section{Derivation of potentials $U_{\phi\xi}^{(1)}$ and $U_{\phi\xi}^{(3)}$ and calculation of frequency change $\Delta\omega_{10,\mu}$}
\label{app:U1U3}

The starting point to derive the formulas for $U_{\phi\xi}^{(1)}$ and $U_{\phi\xi}^{(3)}$ [\esref{U1} and \rref{U3}, respectively] is the following approximation
\be
\sum_m \sin \left[ \sum_\mu W_{\mu m}\xi_\mu \right] \approx -\frac16\! \sum_{m,\mu, \nu, \sigma} \! W_{\mu m} W_{\nu m} W_{\sigma m} \xi_\mu \xi_\nu \xi_\sigma \, ,
\ee
where we used the property $\sum_m W_{\mu m} = 0$ to eliminate the lowest order contribution. Next, we express the collective mode coordinates in terms of creation/annihilation operators and after normal ordering we find
\be\label{sinca}\begin{split}
\sum_{\mu, \nu, \sigma}  W_{\mu m} W_{\nu m} W_{\sigma m} \xi_\mu \xi_\nu \xi_\sigma = \sum_{\mu, \nu, \sigma}  W_{\mu m} W_{\nu m} W_{\sigma m} \ell_\mu \ell_\nu \ell_\sigma
\\ \frac{1}{2\sqrt{2}}\left[\left(a^\dagger_\mu a^\dagger_\nu a^\dagger_\sigma + 3 a^\dagger_\mu a^\dagger_\nu a_\sigma+ \mathrm{H.c.}\right) +3 \left(a^\dagger_\mu+ a_\mu\right)\delta_{\nu\sigma} \right],
\end{split}\ee
where H.c. denotes the Hermitian conjugate.
The last term in square brackets gives the linear interaction term $U_{\phi\xi}^{(1)}$:
\be\label{U11}
U_{\phi\xi}^{(1)} = -\frac12 E_J^a \sin \frac{\phi}{N} \sum_m \sum_{\mu, \nu} W_{\mu m} W_{\nu m}^2 \frac{\ell_\nu^2}{2} \xi_\mu \, .
\ee
To proceed further, we note that
\be\label{W31}
\sum_m W_{\mu m} W_{\nu m} W_{\sigma m} = 1/\sqrt{2N}
\ee
if one of three conditions is satisfied:
\bea
\mu+\nu-\sigma &=& 0 \, , \\
\mu-\nu+\sigma &=& 0 \, ,\\
\mu-\nu-\sigma &=& 0 \, ,
\eea
while
\be
\sum_m W_{\mu m} W_{\nu m} W_{\sigma m} = -1/\sqrt{2N}
\ee
if
\be\label{W32c}
\mu+\nu+\sigma = 2N \, .
\ee
The sum over $m$ vanishes otherwise.
For the sum in \eref{U11} this only leaves two possibilities, $\mu = 2\nu$ and $\mu = 2(N-\nu)$, which imply that $\mu$ must be even. Setting $\mu = 2\rho$ and using the definitions of $\eta_\rho$ and $N_e$ given after \eref{Leven}, we can finally write \eref{U11} in the form given in \eref{U1}.

Using again the identities in \esref{W31}-\rref{W32c}, it is straightforward to cast the terms with products of three operators in \eref{sinca} in the form of \eref{U3}. Here we focus on the calculation of $\Delta\omega_{10,\mu}$, the change in frequency due to the interactions in $U^{(3)}_{\phi_\xi}$, \eref{U3}, when a single collective mode is excited. To this end, let us define the corrections $\delta \epsilon_{l,0}$ and $\delta\epsilon_{l,\bar\mu}$ to the qubit energy $\epsilon_l$ depending on whether the collective modes are in their ground state or in state $|1_{\bar\mu}\rangle$, respectively. These corrections can be calculated at second order in perturbation theory in $U_{\phi\xi}^{(3)}$, for example:
\be
\delta\epsilon_{l,0} = \sum_{j,f}\frac{\left|\langle j,f|U_{\phi\xi}^{(3)}|l,0 \rangle\right|^2}{\epsilon_l - \epsilon_j - E_f} \, .
\ee
Here $|j,f\rangle$ denotes a generic state $|j\rangle$ for the qubit, with energy $\epsilon_j$, and some Fock state $|f\rangle$ with energy $E_f$ for the collective modes.
The similar expression for $\epsilon_{l,\bar\mu}$ is
\be\label{delm}
\delta\epsilon_{l,\bar\mu} = \sum_{j,f}\frac{\left|\langle j,f|U_{\phi\xi}^{(3)}|l,1_{\bar\mu} \rangle\right|^2}{\epsilon_l +\omega_{\bar\mu} - \epsilon_j - E_f} \, .
\ee
In terms of these corrections the qubit frequency change is
\be
\Delta\omega_{10,\bar\mu} = \left(\delta\epsilon_{1,\bar\mu} - \delta\epsilon_{0,\bar\mu}\right) - \left(\delta\epsilon_{1,0} - \delta\epsilon_{0,0}\right) \, .
\ee
A great simplification in calculating $\Delta\omega_{10,\bar\mu}$ is achieved by noticing that in the difference $\delta\epsilon_{l,\bar\mu} - \delta\epsilon_{l,0}$ all contributions originating from terms in $U_{\phi\xi}^{(3)}$, \eref{U3}, for which
none of the indices of the operators in that equation coincides with $\bar\mu$ cancel out. In other words, only terms for which at least one index is $\bar\mu$ can contribute to the frequency change. Moreover, since the Hermitian conjugate terms in \eref{U3} contain two or more annihilation operators, they also do not contribute to $\Delta\omega_{10,\bar\mu}$.
To concretely calculate the matrix elements entering \eref{delm}, we need to consider the action of the terms in square brackets in \eref{U3} onto the singly excited state $|1_{\bar\mu}\rangle$; for example, the first one gives:
\bea\label{u31mu}
\sum_{\mu,\nu} && a^\dagger_{\mu+\nu}a^\dagger_\mu a^\dagger_\nu |1_{\bar\mu}\rangle = \sum_{\mu,\nu}{}^{'} |1_{\bar\mu} 1_\mu 1_\nu 1_{\mu+\nu} \rangle \\ && + 2 \sum_\nu{}^{'} \sqrt{2} | 2_{\bar\mu} 1_{\bar\mu+\nu} 1_{\nu} \rangle + \sum_\nu{}^{'}\sqrt{2} | 2_{\bar\mu} 1_{\bar\mu-\nu} 1_{\nu} \rangle \nonumber \\ &&
+\sqrt{6} |3_{\bar\mu} 1_{2\bar\mu} \rangle + \left(\sqrt{2}\right)^2 | 2_{\bar\mu} 2_{\bar\mu/2} \rangle  \, ,\nonumber
\eea
where the prime at the summation symbols implies that all the collective mode indices in the states being summed must be different. The first term on the right hand side is the one in which no index coincides with $\bar\mu$ and, as discussed above, this term does not contribute to the difference $\delta\epsilon_{l,\bar\mu} - \delta\epsilon_{l,0}$. The last two terms arise from particular combination of the indices ($\mu=\nu=\bar\mu$ and $\mu=\nu=\bar\mu/2$, respectively); the last one is present only if $\bar\mu$ is even. Similar terms in which two indices are equal also arise from the other operators in square brackets in \eref{U3} \cite{fthree}. However, we discard these terms in comparison with the terms with sum over index $\nu$ (see the second line in \eref{u31mu}): because of the sums, the discarded terms are smaller by a factor of order $1/N$.
Taking these considerations into accounts, standard calculation of the matrix elements for the collective modes gives
\begin{widetext}
\bea\label{ded}
\delta\epsilon_{l,\bar\mu} - \delta\epsilon_{l,0} & \approx & \frac{\left(E_J^a\right)^2}{16N} \sum_{j} \left|\langle j | \sin \frac{\phi}{N}| l\rangle \right|^2 \ell_{\bar\mu}^2
\bigg[ \sum_{\nu=1}^{N-\bar\mu-1}\ell_\nu^2\ell_{\bar\mu+\nu}^2\left( \frac{1}{\epsilon_l - \epsilon_j -\omega_{\bar\mu} - \omega_\nu - \omega_{\bar\mu+\nu}} + \frac{1}{\epsilon_l - \epsilon_j +\omega_{\bar\mu} - \omega_\nu - \omega_{\bar\mu+\nu}} \right) \nonumber \\ && + \frac14
\sum_{\nu=1}^{\bar\mu-1}\ell_\nu^2\ell_{\bar\mu-\nu}^2\left( \frac{1}{\epsilon_l - \epsilon_j -\omega_{\bar\mu} - \omega_\nu - \omega_{\bar\mu-\nu}} + \frac{1}{\epsilon_l - \epsilon_j +\omega_{\bar\mu} - \omega_\nu - \omega_{\bar\mu-\nu}} \right) \\ && + \frac14
\sum_{\nu=N-\bar\mu+1}^{N-1}\ell_\nu^2\ell_{2N-\bar\mu-\nu}^2\left( \frac{1}{\epsilon_l - \epsilon_j -\omega_{\bar\mu} - \omega_\nu - \omega_{2N - \bar\mu - \nu}} + \frac{1}{\epsilon_l - \epsilon_j +\omega_{\bar\mu} - \omega_\nu - \omega_{2N - \bar\mu - \nu}} \right) \nonumber
\bigg],
\eea
\end{widetext}
where the approximate equality indicates that we are neglecting $1/N$ corrections originating from terms like the last two in \eref{u31mu}.

To estimate a bound on the frequency change $\Delta\omega_{10,\bar\mu}$, we note that for $l=0, \, 1$ so long as $\omega_{10} < 2\omega_1 - \omega_{N-1}$ there are no divergent contributions in \eref{ded} and the second terms in round brackets are larger than the first ones. For an order-of-magnitude estimate, we further neglect the dependence of the oscillator lengths $\ell_\mu$ and collective mode frequencies $\omega_\mu$ on the array ground capacitance and substitute in \eref{ded} the values $\ell_0$ and $\omega_p^a = \sqrt{8 E_C^a E_J^a}$, respectively. In this way we find (for $l=0, \, 1$)
\bea
&& \left| \delta\epsilon_{l,\bar\mu} - \delta\epsilon_{l,0} \right| \lesssim \frac{E_L}{16} \frac{E_J^a}{N^2} \ell_0^6 \sum_j \left|\langle j | N \sin \frac{\phi}{N}| l\rangle \right|^2  \\ && \qquad \left[ \left( \sum_{\nu=1}^{N-\bar\mu-1} + \frac14 \sum_{\nu=1}^{\bar\mu-1} + \frac14 \sum_{\nu=N-\bar\mu+1}^{N-1} \right) \frac{2}{\omega_p^a + (\epsilon_j - \epsilon_l)}\right]. \nonumber
\eea
With the performed approximations, there is no dependence on collective mode indices, so that the sums in round brackets can be bounded by $N$. An upper bound for the term in square brackets is then $2N/\omega_p^a$ \cite{fub}.
Finally we approximate the qubit matrix element as follows:
\be
\sum_j \left|\langle j | N \sin \frac{\phi}{N}| l\rangle \right|^2 \approx \left | \langle l | \phi^2 | l \rangle \right|\, .
\ee
Rough upper bounds for the latter matrix elements are $(2\pi)^2$ for $l=1$ and $\pi^2$ for $l=0$, see \eref{pmeb}.
We thus find that the bound is tighter for the $l=0$ correction compared to the $l=1$ one, and hence
\be
\left| \Delta\omega_{10,\bar\mu} \right| \le 2 \left| \delta\epsilon_{l,\bar\mu} - \delta\epsilon_{l,0} \right|  \lesssim \frac{\pi^2}{2} \frac{E_L}{N} \frac{E_J^a}{\omega_p^a} \ell_0^6 \, .
\ee

We now compare the frequency change $\Delta\omega_{10,\bar\mu}$ with that due to the dispersive shift $\chi^{\delta E_L}_\mu$ originating from the quadratic interaction $U^{(2)}_{\phi\xi}$, see \eref{chidel}; the latter frequency change is given, in order of magnitude, by the upper bound in \eref{chidelb}, so that an approximate bound on the ratio between the two quantities is
\be\label{dochidelc}
\frac{\left| \Delta\omega_{10,\bar\mu} \right|}{2\chi^{\delta E_L}_\mu} \lesssim (\pi \ell_0)^2 \frac{E_L}{\omega_{10}} \, ,
\ee
where we used $E_J^a/\omega_p^a = \ell_0^{-2}$. For typical experimental parameters, the right hand side is at most of order unity for any value of flux. Given that our approximations place a loose bound on $\Delta\omega_{10,\bar\mu}$, we conclude that the latter can generally be neglected in comparison with $\chi^{\delta E_L}_\mu$.

\section{Derivation of \eref{hcs}}
\label{app:hcs}

For the circuit considered in Sec.~\ref{sec:sic} (symmetrically placed coupling capacitors and no ground capacitance) the matrix $G$ entering \eref{TG} takes the form:
\bea
G_{00} & =  & \frac{1}{2E_C^c} \left(1-\frac{2\delta}{N}\right)^2 , \label{G00c} \\
G_{0\mu} & = & -\frac{1}{E_C^c} \left(1-\frac{2\delta}{N}\right) \frac{1}{\sqrt{2N}} \frac{s_{2\mu\delta}}{s_\mu} o_{\mu+1} \, ,\label{G0mc}\\
G_{\mu\nu} & = & \frac{1}{E_C^c} \frac{1}{N} \frac{s_{2\mu\delta}}{s_\mu} \frac{s_{2\nu\delta}}{s_\nu} o_{\mu+1} o_{\nu+1} \, .\label{Gmnc}
\eea
Note that both matrix $G$ and ${\cal L}_V$ in \eref{LV2} vanish for $\delta = N/2$; this choice (possible only for even $N$) connects both capacitors to the same island, and the vanishing is due to the arbitrariness in choosing a reference for the electric potential. For generic $ 0 < \delta <N/2$ we see that, beside a renormalization of the qubit charging energy [\eref{G00c}], the coupling capacitors couple the qubit and even modes [\eref{G0mc}], similarly to the effect of ground capacitances $C_g^a$ in the array, while at variance with the effect of the ground capacitances, the coupling capacitors lead to mode-mode interaction in the even sector [\eref{Gmnc}], rather than in the odd one. (Of course parity symmetry still guarantees the qubit-odd modes decoupling.) The situation, however, is only apparently complicated, since a change of variables shows that the problem reduces to a single collective mode coupled to the qubit and the cavity, while all other modes remain degenerate and decoupled. Indeed, let us introduce the new even-sector variables:
\bea
\tilde{\eta}_1 &=& \sum_{\rho=1}^{N_e} \frac{v_\rho}{|v|} \eta_\rho \, , \\
\tilde{\eta}_\rho &=& \frac{|v|_{\rho-1}}{|v|_{\rho}} \eta_\rho - \sum_{\sigma=1}^{\rho-1} \frac{v_\rho v_\sigma}{|v|_\rho |v|_{\rho-1}} \eta_\sigma \, .
\eea
This transformation is a rotation for any vector $v$ with the definition
\be
|v|_{\rho}^2=\sum_{\sigma=1}^{\rho} v_\sigma^2
\ee
for the norms, and $|v| \equiv |v|_{N_e}$. In the present case we take the components $v_\rho$ to be
\be
v_\rho = \frac{1}{\sqrt{N}} \frac{s_{4\rho \delta}}{s_{2\rho}}
\ee
and find that the kinetic energy term $T_G$ simplifies to
\be\label{TGcs}\begin{split}
T_G  =  & \frac{1}{16}\bigg[G_{00} \dot\phi^2  + \frac{|v|^2}{E_C^c} \dot{\tilde{\eta}}_1^2  - 2\frac{1-2\delta/N}{\sqrt{2}E_C^c}|v|\dot{\tilde{\eta}}_1 \dot\phi\bigg].
\end{split}\ee
Similarly, ${\cal L}_V$ of \eref{LV2} becomes
\be\label{LVcs}
{\cal L}_V = -\frac{1}{4E_C^c} \left(1-\frac{2\delta}{N}\right) \dot\phi eV + \frac{1}{2\sqrt{2} E_C^c} |v| \dot{\tilde{\eta}}_1 eV \, .
\ee
The norm of vector $v$ can be calculated explicitly \cite{trigsum}:
\be
|v|^2 = \delta \left(1-\frac{2\delta}{N}\right) \, .
\ee

Neglecting the array non-linearities, the total Lagrangian is the sum of ${\cal L}_U$ in \eref{LU} with $T_G$ and ${\cal L}_V$ of \esref{TGcs}-\rref{LVcs}. Then performing the Legendre transform and defining
\be
\xi_\mu = \left\{\begin{array}{ll}
\tilde{\eta}_{\mu/2}\, , & \mu \ \mathrm{even} \\
\zeta_{(\mu+1)/2} \, , & \mu \ \mathrm{odd}
\end{array}\right.
\ee
for $\mu = 1,\ldots,N-1$, we arrive at the Hamiltonian in \eref{hcs}.


\begin{thebibliography}{30}
\bibitem{lappr}As explained before \eref{U}, the approximation consists in neglecting the array junction non-linearities. We will consider their effects and the resulting interactions between even and odd sectors in Sec.~\ref{sec:anhar}. Here we note that the non-linearities preserve $PT$-symmetry and therefore cannot introduce even/odd couplings at the quadratic level.
\bibitem{fpert}Modes with index $\sigma \gg \sigma^* = (N/2)\sqrt{E_C^a/4E_g^a}$ can always be treated perturbatively; for parameters in Table~\ref{tabpar} we find $\sigma^*\approx 1$, so most modes, except the few lowest ones, are in fact in the perturbative regime.
\bibitem{flow}We consider low occupation of the collective modes, in which case this condition agrees with the more precise one discussed in \ocite{fluxth}.
\bibitem{fid}Identities with the structure of \eref{meid} are generally valid for Hamiltonians with quadratic kinetic energy terms, as can be shown by calculating
{$\left| \langle m | [H,\phi] | l \rangle \right|^2$} either by evaluating the commutator first, $[H,\phi] \propto p_\phi$ , or by applying the Hamiltonian to the eigenstates.
\bibitem{notechiflux}The exceptions are narrow regions in flux around those points where the shift $\chi_\rho$ has zeros.
\bibitem{fdeph}We note that in fact \eref{Garho} is more generally valid for $\bar{n}_\mu \ll \max\{1, 2\bar\chi_\mu/\kappa_\mu\}$.
\bibitem{fodd}The odd modes being decoupled, their decay rate is formally zero. As we show in Sec.~\ref{sec:sb}, unavoidable deviation from perfect parity symmetry will introduce coupling between the odd mods to the cavity, but this is usually negligible.
\bibitem{fasy}Asymmetry in the ground capacitances $C_g^b$ next to the phase-slip junction can be accounted for by the replacements $\delta c \to \delta c + \delta c_b$ in \esref{G00s}-\rref{G0ms}, with $\delta c_b = (C_g^0 - C_g^N)/C_t$, as well as $(\delta c)^2 \to \delta c (\delta c + \delta c_b)$ in the second term of \eref{LVs}.
\bibitem{ffail}This relationship will eventually fail as $N$ increases to saturate the bound in \eref{Hcon}.
\bibitem{fthree}For the last two terms in square brackets in \eref{U3}, if $N$ is multiple of 3 then for $\bar\mu=2N/3$ all indices can be equal.
\bibitem{fub}Strictly speaking, for $l=1$ the upper bound for the term with $j=0$ in the sum over $j$ is $2N/(\omega_p^a - \omega_{10})$. However we note that even at zero external flux, the qubit frequency is in practice sufficiently smaller than the collective mode frequency that our neglecting $\omega_{10}$ is justified for an order of magnitude estimate. Moreover, when comparing to $\chi^{\delta E_L}_\mu$ in \eref{dochidelc}, the possible underestimation of $\Delta\omega_{10,\bar\mu}$ due to this approximation is compensated for by the prefactor $E_L/\omega_{10}$, which is minimum at zero external flux.
\end{thebibliography}
\end{document}